\documentclass[superscriptaddress,floatfix,showkeys,showpacs]{revtex4}

\usepackage{amscd}
\usepackage{epsfig}
\usepackage{amsmath}
\usepackage{amssymb}
\usepackage{epsf}

\newtheorem{definition}{Definition}

\newtheorem{theorem}{Theorem}
\newtheorem{corollary}{Corollary}

\newtheorem{conjecture}{Conjecture}

\begin{document}
\bibliographystyle{plain}


\title{Probability of local bifurcation type from a fixed point: A random matrix perspective}
\author{D. J. Albers}
\email{albers@cse.ucdavis.edu}
\affiliation{Max Planck Institute for Mathematics in the Science, Leipzig 04103, Germany}

\author{J. C. Sprott}
\email{sprott@physics.wisc.edu}
\affiliation{Physics Department, University of Wisconsin, Madison, WI 53706}

\date{\today}


\begin{abstract}
Results regarding probable bifurcations from fixed points are
presented in the context of general dynamical systems (real, random
matrices), time-delay dynamical systems (companion matrices), and
a set of mappings known for their properties as universal approximators
(neural networks).  The eigenvalue spectra is considered both
numerically and analytically using previous work of Edelman et. al.
Based upon the numerical evidence, various conjectures are presented.
The conclusion is that in many circumstances, most bifurcations from
fixed points of large dynamical systems
will be due to complex eigenvalues.  Nevertheless, surprising
situations are presented for which the aforementioned conclusion is
not general, e.g. real random matrices with Gaussian elements with a
large positive mean and finite variance.

\end{abstract}


\maketitle




\section{Introduction}

Determination of the types of random matrices that constitute a
physically relevant set depends markedly on the field of study.  From the perspective of those interested in quantum mechanical
phenomena (e.g. nuclear physics), one might be led to believe that
random matrices that are not unitary or Hermitian are of no particular
physical interest \cite{mehta_bookz_chapter}\footnote{In chapter $15$
  of Mehta's famous book one can find the words ``An ensemble of
  matrices whose elements are complex, quaternion, or real numbers, but
with no other restrictions as to their Hermitian or unitary character,
is of  no immediate physical interest, for their eigenvalues may lie
anywhere on the complex plane.''  He later asserts that these matrices
are, nevertheless, of interest in their own right.}.  However, if one were interested in dynamics a la Poincar\'e, where an
understanding of the derivative along an orbit is of utmost importance and where
Lyapnov exponents are the key quantities of interest; then matrices of prime concern are real random
matrices from the general linear group (the dissipative case) or the
special linear group (the non-dissipative case).  The interest in real
random matrices arises because they form the
linear portion of the derivative or the tangent space of a dynamical
system at a given point
along a trajectory \cite{ams_rmt} \cite{ruellehilbert} \cite{katok80}
\cite{oseledec}, \cite{pesinmet3}.  From a dynamical reconstruction perspective,
where time-series data is used \cite{deelepaper}, it is companion
matrices and thus polynomials with random coefficients that are of
interest.  Because we are concerned with dynamical systems in general,
specifically probabilities of bifurcations from fixed points and the
factors that determine those
probabilities, we will focus on
real random matrices with various distributions followed by a practical
construction using ``universal approximators'' with random weights.  

To place the current study in context, consider first the following
background.  Doyon et. al. \cite{doy} argued that the most likely first bifurcation and route
to chaos given a particular set of dynamical systems was that of the Ruelle-Takens quasi-periodic
route to chaos based upon the random matrix results of
Girko \cite{girko1} (for other useful versions of this random matrix
result see Edelman \cite{edelman1}, and Bai \cite{bai}).  Likewise,
Sompolinsky et. al. \cite{Sompolinsky_nn} showed that as the dimension
of a dynamical system is increased, the location of the bifurcation from a fixed point
decreased toward zero; and the routes to chaos region of parameter space subject to parameter
variation decreased in length.  Doyon et. al. and Sompolinsky
et. al. both considered neural networks with a single hidden layer
(a single layer of neurons) whose input layer was entirely replaced by
it's output layer at each time-step (i.e. ``vector'' neural networks).  However, Sompolinsky
et. al. considered the continuous time case where as Doyon
et. al. worked in discrete time --- both constructions yielded a
similar set of conclusions.  In previous publications \cite{albersroutetochaosI} the authors
considered time-delay neural networks such as those presented in the
work of Hornik et. al. \cite{hor2} that have been found to be universal
approximators (i.e. they can approximate $C^r$ ($r \geq 0$ mappings and
their derivatives on compact, metrizible sets) and also came to similar conclusions despite the
significant differences in the constructions.  In \cite{albersroutetochaosII}
the authors claimed that as the dimension of the dynamical system is
increased, flip, fold, or any bifurcations due to real eigenvalues, and
strong resonance bifurcations will be vanishingly rare and the route
to chaos from a fixed point in parameter space will consist of
periodic orbits with high-period ($>4$) and quasi-periodic orbits.
The basic idea of the above arguments follows from the matrix theory of Girko \cite{girko1}, Edelman
\cite{edelman1}, and Bai \cite{bai} and is the following: given a square matrix whose
elements are real random variables drawn from a distribution with a
finite sixth moment, in the limit of infinite dimensions, the
normalized spectrum (or eigenvalues) of the matrix will converge to a uniform
distribution on the unit disk in the complex plane.  It is worth
noting that the convergence in measure is not absolutely continuous with respect to
Lebesgue measure.  Nevertheless, if the Jacobian of the map at the ``first'' bifurcation point
(i.e. the bifurcation from the fixed point) is a high-dimensional matrix
whose elements have a finite sixth moment, it is reasonable that the
bifurcation would be of type Neimark-Sacker (via a complex
eigenvalue) instead of a
flip or fold (via a real eigenvalue), with probability approaching unity as the dimension goes
to infinity.  


Each of the studies mentioned
above consist of a statistical sampling of a space  of mappings via a
weight structure imposed upon those mappings.  In a sense, it is a
statistical sampling of the effects that the weight matrices have on
the dynamics.  The evidence discussed in the
former paragraph leads to at least three important questions: how
robust are the results with respect to the measures imposed upon the
weight matrices; given that there do exist observable period-doubling
bifurcations in high-dimensional dynamical systems, how does this
occur given the random matrix style arguments; how can these results
be connected and compared with real world systems --- what are the
links with the natural world?  Addressing the first two questions is a
matter of carefully studying how the results from random matrix theory
apply specifically to what has been observed computationally.
Discussion of the third issue lies with providing a construction that
yields comparison with real-world data.  

As we will see, the distributions and perturbations of the
distributions of the elements of random matrices can have profound
and surprising effects on a bifurcation sequence while having negligible effects
regarding the generality of the proven theorems.  Thus, perturbations
of the weight distributions of the computationally studied systems
can be made in such a way that  significantly alters the conclusions of the random
matrix arguments as applied to bifurcation theory.  This will provide
answers to why and how period-doubling sequences can be observed in
high-dimensional systems.  This of course does not marginalize the
results discussed above; it only qualifies them, for despite the
effects we observe, the former results remain quite general.  

We will begin with a general study of linear dynamical systems at fixed points.  Generically, all dynamical systems at stable fixed
points can be thought of as linear maps via the implicit function theorem.
We will identify the Jacobian of the aforementioned dynamical systems
with random matrices with various distributions.  We will follow this approach, adopting the framework of random
polynomials which can be likened to companion matrices.  The linear
part of the derivative of time-delay dynamical systems is that of
a companion matrix.  Thus, we will study general, linear, time-delay
dynamical systems subject to weight distributions of the coefficients
of the characteristic polynomial.  The final framework we will use --- which was begun in
\cite{albersroutetochaosI} --- is that of scalar, time-delay neural
networks that are commonly used to reconstruct a dynamical system from
empirical, time-series data.  This will begin to forge a connection
with natural systems because networks such as the ones we study can be
used to fit other data sets, and the weight distributions of the
trained networks can be compared with those from more theoretical distributions.

\section{Construction}
\label{sec:construction}
Begin with a discrete-time, $C^r$ ($r>0$, $r \in N$) dynamical system that maps a compact subset
$U \subset R^d$ to itself:
\begin{equation}
\label{eqn:ds1}
F: U \rightarrow U
\end{equation}
However, due to practical motivations we will instead consider a set
of functions that can universally approximate a properly chosen time-delay map of
$F$ (we will discuss this more in section \ref{sec:approximation}).  Specifically, we consider  $f$ to be explicitly defined:
\begin{equation}
y_{t+1} = f(y_t, \dots, y_{t-d})
\end{equation}
where $f$ is $C^r$ and $y_i \in R$.  In a practical sense, $f$ can be
thought of as a standard feedforward neural network given by:
\begin{equation}
\label{equation:intro_net}
 x_{t} = \beta_0 + \sum_{i=1}^{n}{{\beta}_i \tanh \left( s{\omega}_{i0} + s
 \sum_{j=1}^{d}{{\omega}_{ij} x_{t-j} } \right)} 
\end{equation}
which is a map from $R^{d}$ to $R$. Here $n$ is the number of hidden units
(neurons), $d$ is the number of time lags which determines the system's input
(embedding) dimension, and $s$ is a scaling factor on the connection weights
$w_{ij}$. The initial condition is $({x}_{1}, x_{2}, \ldots, x_{d})$ and the
state at time $t$ is $({x}_{t}, x_{t+1}, \ldots, x_{t+d-1})$. The approximation
theorems of Ref. \cite{hor2} and well known time-series embedding results
Ref. \cite{takensstit} together establish an equivalence (given
properly chosen weight distributions) between this class
of neural networks and the general dynamical systems of interest here
\cite{hypviolation}.

We sample the $(n(d+1)+1)$-dimensional parameter space taking (i)
$\beta_{i} \in [0,1]$ uniformly distributed and rescaling them
to satisfy $\sum_{i=1}^{n}{{{\beta}_{i}^{2}}} = n$, (ii) ${w}_{ij}$ as
normally distributed with zero mean and unit variance, and (iii) the initial
$x_j \in [-1,1]$ as uniform.  We will focus largely on
behavior types as a function of the parameter $s$, which can be interpreted
as the standard deviation of the $w$ weight matrix, and the embedding
dimension $d$.  Note that for $x \approx 0$ $\tanh(x)$ is nearly linear. Thus,
choosing $s$ to be small forces the dynamics to be mostly linear,
yielding fixed-point behavior. Increasing $s$ yields a route to chaos
\cite{albersroutetochaosI} \cite{albersroutetochaosII}.  Due to this, $s$
provides a unique bifurcation parameter that sweeps from linear to highly
nonlinear parameter regimes.  In this paper we will however only
consider the networks up to the first bifurcation.

Time-delay, scalar networks have an extremely convenient property that aids in the
calculation of the local derivative and in the computation of the
eigenvalue of the Jacobian at a fixed point; the local derivative of our scalar networks are
companion matrices:

\[ Df_x = \left[ \begin{array}{ccccccc}
      a_{1} & a_2 & a_3 & \cdots & a_{d-2} & a_{d-1} & a_{d} \\
      1 & 0 & 0         & \cdots & 0 & 0 & 0 \\
      0 & 1 & 0 & \cdots         & 0 & 0 & 0 \\
      \vdots & \ddots & \vdots \\
      0 & 0 &0 &\cdots & 0 & 1 & 0 \\

\end{array} 
\right]\] 

The $a_k$'s are given as:
\begin{equation}
\label{eqn:aks}
a_k = \frac{\partial x_{t}}{\partial x_{t-k}} = \sum_{i=1}^{n}
\beta_i s w_{i k} sech^2( s w_{i 0} + s \sum_{j=1}^d w_{ij} x_{t-j}) 
\end{equation}    


%


\subsection{Approximation theory}
\label{sec:approximation}

To understand how the neural networks in our study fit into the
dynamical systems framework, we must consider two issues, the
relationship between time-delay dynamical systems and the
approximation capabilities of neural networks.


The answer to the
first issue is given by Takens --- and
more tailored to our construction, the paper of Sauer
et. al. \cite{embedology} --- is yes with certain constraints on
$F$.  Namely, the box-counting dimension of the set of periodic orbits
of period $p$, ($A_p$) must be less than $\frac{p}{2}$, and $DF_p$
must have distinct eigenvalues.  These constraints are likely
satisfied since we do not consider portions of parameter space that
easily yield periodic orbits.  

\begin{figure}
\begin{center}
\epsfig{file=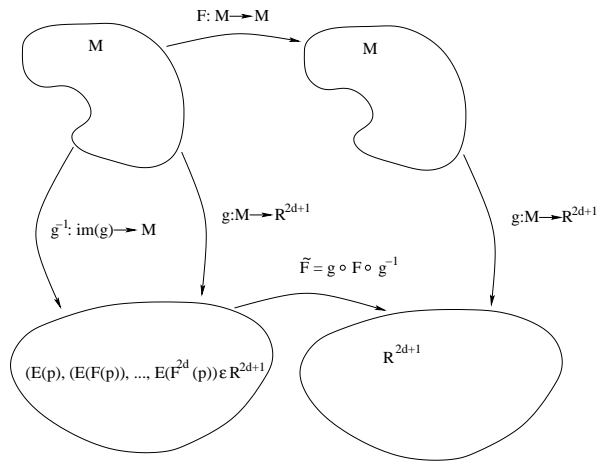, height=6cm}
\caption{Schematic diagram of the Takens embedding theorem and how it
  applies to our construction.}
\label{fig:takensfigure}
\end{center}
\end{figure}

The relationship of general time-delay dynamical systems is given in Fig.
(\ref{fig:takensfigure}) in which $F$ is a $C^r$ dynamical system, $E: M \rightarrow R$ a is a
``measurement function,'' ($E$ is a $C^k$ map) embedding $g:M \rightarrow R^{2d+1}$ is explicitly given by:
\begin{equation}
g(x_t) = (E(x_t), E(F(x_t)), \dots, E(F^{2d}(x_t)))
\end{equation} 
In a colloquial, experimental sense, $\tilde{F}$ just keeps track of
the observations from the measurement function $E$, and, at each time
step, shifts the newest observation into the $2d+1$ tuple and
sequentially shifts the scalar observation at time $t$ ($y_t$) of the $2d+1$ tuple to the
$t-1$ position of the $2d+1$ tuple.  In more explicit notation, $\tilde{F}$ is the following mapping:
\begin{equation}
\label{equation:F-tilde}
(y_1, \dots, y_{2d+1}) \mapsto (y_2, \dots, y_{2d+1}, g(F(g^{-1}(y_1,
  \dots, y_{2d+1}))))
\end{equation}
where, again, $\tilde{F} = g \circ F \circ g^{-1}$.

The response to the question regarding the approximation capabilities of neural
networks is significantly more
complicated due to two particular issues ---  the networks
have only finitely many parameters and the induced measure on the weights of the neural networks determines
what phenomena will be fit stably.  That networks we consider have finitely many parameters implies there will
never exist a one-to-one or onto correspondence between the
space of $C^r$ diffeomorphisms and neural networks.  The best that can
be achieved is a density in the relevant function space with
infinitely many parameters.  This is a fundamental functional approximation issue that has no
chance of being improved, however, this issue is clearly not terminal
for our study.  The more interesting problem is that of the induced measure ---
specifically, how the space of neural networks with the induced measure on the parameter space represent
approximates of the dynamical systems that satisfy the technical
restrictions of Sauer et. al.  Further, what subsets of the space
of $C^r$ dynamical systems are being selected out by such a measure
(or even our measure) is a very interesting question.  There surely exists a measure on the space of
parameters of neural networks such that (given the dimension is high
enough) the selected neural networks will
approximate and correspond to a set of the $C^r$ mappings that satisfy the
restrictions of Sauer et. al.  In fact our measure often satisfies
this criterion given certain restrictions.  Such issues hopefully can be addressed by applying information
geometry to the parameter space of neural networks following work of
\cite{amaribook}.  However, a clear understanding of what portion of the $C^r$ mappings we
are sampling is currently difficult to discern.

Setting the issues of a finite number of parameters aside and leaving
any constraints on the weights behind, the neural
networks we utilize can
approximate $\tilde{F}$ and its derivatives (to any order) to
arbitrary accuracy \cite{hor} \cite{hor2}.  Precise statements of the
neural network approximation theorems require machinery from
functional analysis (see \cite{adams}) and is covered in detail in the
papers by Hornik et. al. and is discussed in \cite{hypviolation}.  The
neural networks can approximate mappings from Sobolev spaces, $S^m_p(U, \lambda)$, such as
the one defined here:
\begin{definition}
For any positive integer $m$ and $1 \leq p < \infty$, we define a
Sobolev space $S^m_p(U, \lambda)$ as the vector space on which $|| \cdot
||_{m,p}$ is a norm:
\begin{equation}
S_p^m(U, \lambda) = \{ f \in C^m(U) | \text{  } ||D^{\alpha}f||_{p, U, \lambda}
< \infty \text{ for all } |\alpha| \leq m \}
\end{equation}
Equipped with the Sobolev norm, $S_p^m$ is a Sobolev space over $U
\subset R^d$.
\end{definition}    
Neural networks can also approximate most $C^r$ mappings and their
derivatives.  However, if
the mapping is piecewise differentiable, the neural networks must be
fit independently on both sides of the discontinuity.  Nevertheless,
neural networks form a set of ``universal approximators'' --- they can
approximate nearly any imaginable mapping.


\subsubsection{The implicit measure}
\label{sec:takens_nn_implicitmeasure}

In a very practical sense, we are implicitly imposing a measure
on the set of neural networks via the probability distributions of the
weights on $R^{n(d+2)+1}$ (i.e. the probability distributions form a
product measure on $R^{n(d+2)+1}$).  This will
introduce a bias into our results that is unavoidable in such
experiments; the very act of picking networks out of the space will
determine, to some extent, our results.  Examples of how important
this is will be revealed as we consider the numerical results.  Unlike actual physical experiments, we could, in principle,
prove an invariance of our results to our induced measure, however this is
difficult and beyond the scope of this paper.  It suffices
for our purposes to note specifically what our measure is (the weight
selection method), and how it
might bias our results.  The selection method will include all
possible networks, but clearly not with the same likelihood.  In the absence of a theorem with
respect to an invariance of our induced measure, we must be careful
in stating what our results imply about the ambient function space.

\subsection{Why scalar neural networks?}
\label{sec:why_scalar_nets}

Time-delay scalar neural networks are enticing to study because they
allow for a combination of the perspectives of functional analysis, topological dynamics, and
the practical, real world.  From the functional analysis perspective,
the neural networks we study are diverse universal function
approximators --- they can approximate nearly any mapping one might
wish to approximate \cite{hor2}.  Thus, studying the space of neural
networks represents a practical study of a common space of mappings used by
time-series analysists to reconstruct unknown dynamics from
time-series data \cite{deelepaper} \cite{sprott_book} \cite{kantz_book}.  From the perspective of topological dynamics, if
one wishes to study general dynamical systems, there is always the
issue of how to relate such studies to the natural world that they
were originally meant to model.  Studying time-delay dynamics as they relate to $C^r$ dynamical systems
is a partial connection or link between abstract
dynamics and the natural world since the natural world is often
studied with
time-series data \cite{yorke_observe_embed}.
Yet there is the issue of
relating time-delay dynamics to a specific natural system or class of
systems.  However, one of the key
sources of the problem of relating the space of neural networks to a
space of general dynamical systems is the
implied measure on the weight space.  This is also where the practical link
between the natural world and general time-delay dynamical systems can
be found.  The weight distributions form a clear link between the
abstract dynamics world via embedding and function approximation of
real time-series data from natural phenomena.  If the dynamics that arise from the space of neural
networks can be understood in terms of their weight distributions and
on the neural networks can be fit to data from nature ---
the fit weight distributions can be compared with the dynamics
dependent on the weight distributions.  Thus, aside from being an extremely general class
of dynamical systems and universal approximators in their own right,
neural network weight distributions are a possible link between the
abstract world of mathematical dynamics and the natural world.


\section{Random Matrix Theory}
\label{sec:randommatrixtheory}
Our discussion of random matrix theory will be limited to the circular
law of Girko \cite{girko1}, Bai \cite{bai}, and Edelman
\cite{edelman1}.  In general, circular laws in random matrix theory relate the
distributions of elements of a random matrix to the distribution of
those matrices' eigenvalues on a disk, usually centered at the origin,
in the complex plane.  We will begin by discussing the
circular law outright and follow this with a discussion of the
expected value of real eigenvalues of a random matrix, and various
related results.  In both of the sections that follow, all of our matrices will be $n
\times n$ matrices with real elements drawn from a random distribution
yet to be specified.

%

Dynamical systems at fixed points can be identified with random
matrices via the linear derivative map of the dynamical system $f$ where each of the terms in the matrix is given by a number from a
random distribution.  This is quite general --- given that the range
and domain of the dynamical system have the same dimensions.  Any
generic, discrete-time, $d$-dimensional dynamical system at a fixed point can be recast
as a linear map $g = A x$ where $A$ is a $d \times d$ matrix via the
implicit function theorem.  Thus, studying the spectrum of random matrices is, in a
way, equivalent to studying bifurcations of dynamical systems at non-degenerate fixed points with respect to the variation of a linear
scaling parameter; because the
spectrum of $Df$ will yield the entire geometric structure of $f$.




\subsection{The Circular Law}
\label{sec:circularlaw}
The study of the circular law has a long,  somewhat colorful,
and debated history.  In the early $50$'s it was conjectured that the empirical
spectral distribution (i.e. the distribution of eigenvalues) of $n \times n$ matrices with independent and identically
distributed elements that were normalized by $\frac{1}{\sqrt{n}}$,
converged to a uniform distribution on the unit disk in the complex
plane.  This is what is refered to as the circular law.  In $1965$ Ginbre \cite{ginbre} 
proved this conjecture in the case where the random matrix is complex and has
elements whose real and imaginary parts are independent and normally distributed (i.e. the real and imaginary parts are
independent normals).  V.I. Girko published, in
$1984$ \cite{girko1}, $1994$ \cite{girko2},  and again in $1997$
\cite{girko3} \cite{girko4}, papers proving the circular law for real,
random, Gaussian, matrices.  Girko's circular law states that as $n
\rightarrow \infty$, the distribution of $\frac{\lambda}{\sqrt{n}}$ tends to uniformity on the unit
disk.  This result, which implies that as $n \rightarrow \infty$, the
probability of an eigenvalue being real must go to zero, is a key ingredient towards showing that
local bifurcations from fixed points due to purely real eigenvalues will be unlikely.  It is
this result that limits the kinds of generic local bifurcations from
fixed points we can observe in the infinite-dimensional limit.  One particularly unfortunate
problem with Girko's measure (as well as Edelman's and Bai's) is that it is not absolutely
continuous with respect to Lebesgue measure with increasing but finite
dimension (the infinite-dimensional limit is however absolutely
continuous).  In particular, the probability of an eigenvalue being real with
respect to Edelman is higher
than one might expect for finite dimensions.  Thus, convergence
in distribution to uniformity on the unit disk becomes an issue.
However, Edelman \cite{edelman2} derived a formula
for the expectation value of real eigenvalues in Girko's measure in
finite dimensions (we
will discuss that result in section
\ref{sec:expectedvaluerealeigenvalues}) which we will discuss since it
is very important with respect to our results.  In the
process of deriving this expectation formula, Edelman also proved
Girko's result.  In 1997, Bai \cite{bai} provided an alternate proof
of the
circular law for real random matrices with a significantly weaker
hypotheses than either Edelman or Girko.  Bai's result requires that the elements of the
matrix be from a distribution with only a finite sixth moment.  We will
state Bai's result here since is it is both the simplest, and most general.
\begin{theorem}[Circular law \cite{bai} (page 496)]
Suppose that the entries of a $n \times n$ matrix $M$ have finite sixth moment and that the
joint distribution of the real and imaginary part of the entries has a
bounded density.  Then, with probability $1$, the empirical
distribution $\mu_n(x, y)$ tends to the uniform distribution over the
unit disk in two-dimensional space.
\end{theorem}  
Understanding the difference between the results of Bai, Girko,
Edelman, the convergence of the density of eigenvalues on the unit
disc, and how this is related to the eigenvalue with the largest
magnitude will be the focus of sections \ref{sec:m_gaussian} and \ref{sec:m_uniform}.

\subsection{Expected Value of Real Eigenvalues and Related Results}
\label{sec:expectedvaluerealeigenvalues}
The circular law will provide the intuition for the conjecture we will
state shortly, but it is difficult to use for our purposes,
and provides little practical understanding of how the distribution of eigenvalues
evolves and converges to uniformity as the dimension of the matrix is increased.  Luckily,
Edelman essentially evaluated the integral formula of Girko (in spirit,
at least) and arrived at a formula for the expected number of
real and complex eigenvalues as a function of the dimension of the matrix.
Edelman has proved the following formulas and theorems which will be useful and
relevant for our work: a formula for the density of
real eigenvalues in the complex plane as a function of the
dimension of the matrix; a formula for the density of non-real
eigenvalues on the complex plane as a function of the dimension of the
matrix; a formula for the expectation value of real eigenvalues of a
matrix as a function of the dimension of that matrix; and a
theorem that states that the real eigenvalues converge in distribution
to that of a uniform random variable on $[-1, 1]$ in the limit of an
infinite-dimensional matrix.  For completeness, we will reproduce the
aforementioned results, noting that all the statements that follow are relevant for matrices such that $A \in
R^{n^2}$ where $a_{ij} \in N(0, 1)$.

We will begin with two definitions, the true density of real
eigenvalues of a real random matrix and the probability density of real
eigenvalues of the said random matrix.  These can be found in \cite{edelman2}.

Assume $\lambda$ is a real eigenvalue of a fixed, real $n \times n$
matrix $A$.  The true density of real eigenvalues, or
the expected number of real eigenvalues per unit length can be defined:
\begin{equation}
\rho_n(\lambda) = ( \frac{1}{\sqrt{2 \pi}}
[\frac{\Gamma(n-1,\lambda^2)}{\Gamma(n-1)}] + \frac{|\lambda^{n-1}|e^{-\frac{\lambda^2}{2}}}{\Gamma(\frac{n}{2})2^{\frac{n}{2}}}[\frac{\Gamma(\frac{(n-1)}{2},\frac{\lambda^2}{2})}{\Gamma(\frac{(n-1)}{2})}])
\end{equation}
or, in a different light:
\begin{equation}
\rho_n(x) = \frac{d}{dx} E_A \#_{(-\infty, x)}(A)
\end{equation}
where $\#_{(-\infty, x)}(A) \equiv$ number of real eigenvalues of $A
\leq x$, $E_A$ denotes the expectation value for a random $A$ and
$\Gamma$ is the standard gamma function.  Moreover the probability density of $\lambda_n \in R$, $f_n(\lambda)$ is given by:
\begin{equation}
f_n(\lambda) = \frac{1}{E_n} ( \frac{1}{\sqrt{2 \pi}}
[\frac{\Gamma(n-1,\lambda^2)}{\Gamma(n-1)}] + \frac{|\lambda^{n-1}|e^{-\frac{\lambda^2}{2}}}{\Gamma(\frac{n}{2})2^{\frac{n}{2}}}[\frac{\gamma(\frac{(n-1)}{2},\frac{\lambda^2}{2})}{\Gamma(\frac{(n-1)}{2})}])
\end{equation}
or more simply:
\begin{equation}
f_n(\lambda) = \frac{1}{E_n}\rho_n(\lambda)
\end{equation}
where $E_n$ denotes the expected number of real eigenvalues of the
$n \times n$ random matrix.

Integrating $\rho_n$ along the real line provides the expected number
of real eigenvalues.  Edelman provides several
formulas from such a calculation, the simplest being summarized by
the asymptotic series given in corollary $5.2$ of \cite{edelman2}]:
\begin{equation}
\label{equation:ed_formula}
E_n =
\sqrt{\frac{2n}{\pi}}(1-\frac{3}{8n}-\frac{3}{128n}+\frac{27}{1024n^2}
+ \frac{499}{32768n^4}+O(\frac{1}{n^5}))
\end{equation}
Again, $E_n$ is the expected number for a real, $n \times n$ random
matrix.  The manner in which the convergence in measure is not
absolutely continuous (with respect to Lebesgue measure) is highly
relevant to our results --- the manner in which the convergence is not
absolutely continuous with respect to Lebesgue measure results in an
expected density of real eigenvalues that is higher than other chords
of length $2 \pi$.  For a full discussion, see
\cite{edelman2}.  However, this is not a pathological problem since the
distribution of real eigenvalues on the real line is uniform.  Moreover, for our purposes, the expected value of the real eigenvalues  is not
enough since, if all the real eigenvalues are located at
$\pm 1$, then clearly there will exist many local bifurcations from
fixed points due to purely real eigenvalues --- and these bifurcations
will likely be of high codimension.  That such is
not the case is given in the following corollary:
\begin{corollary}[Corollary $4.5$ \cite{edelman2}]
\label{corollary:fourpointfive}
If $\lambda_n$ denotes a real eigenvalue of an $n \times n$ random
matrix, then as $n \rightarrow \infty$, the normalized eigenvalue
$\frac{\lambda_n}{\sqrt{n}}$ converges in distribution to a random
variable uniformly distributed on the interval $[ -1, 1 ]$
\end{corollary}

Besides the results regarding the real eigenvalues, Edelman also
provides information regarding the density of non-real eigenvalues:
\begin{theorem}[Density of Non-Real Eigenvalues: Theorem $6.2$ \cite{edelman1}]
\label{theorem:complexdensity}
The density of a random complex eigenvalue of a normally distributed
matrix is:
\begin{equation}
\rho_n(x, y) = \sqrt{\frac{2}{\pi}}ye^{y^2-x^2} \mathit{erfc}(y\sqrt{2})e_{n-2}(x^2+y^2)
\end{equation}
where $e_n(z)=\sum_{k=0}^n \frac{z^k}{k!}$ and $\mathit{erfc}(z) =
2/\pi \int_z^{\infty} exp-t^2) dt$, the complementary error function.
Integrating this over the
upper half plane gives the number of non-real eigenvalues.
\end{theorem}

All of these results can be nicely concluded with the following two
theorems regarding the circular law.

\begin{theorem}[Theorem $6.3$ \cite{edelman1}]
\label{theorem:63}
The density function $\hat{\rho}$ converges pointwise to a very simple
form as $n \rightarrow \infty$:
\begin{equation}
\lim_{n \rightarrow \infty} \frac{1}{n} \hat{\rho}(\hat{x}, \hat{y}) = 
\begin{cases}
        \frac{1}{\pi} & \hat{x}^2 + \hat{y^2} < 1 \\
        0 & \hat{x}^2 + \hat{y^2} > 1
\end{cases}
\end{equation}
where $\hat{\rho}_n$ is simply $\rho$ as a function of
$\hat{x}=\frac{x}{\sqrt{n}}$ and $\hat{y}=\frac{y}{\sqrt{n}}$.  Note
that $\frac{\hat{\rho}(\hat{x}, \hat{y})}{n}$ 
is a randomly chosen normalized eigenvalue in the upper half plane.
\end{theorem}

Finally, Edelman's version of the circular law can be proved using theorem \ref{theorem:63} and a central
limit theorem.
\begin{theorem}[Circular Law: Convergence in distribution \cite{edelman1}]
\label{theorem:circlaw}
Let $z$ denote a random eigenvalue of $A$ chosen with probability
$\frac{1}{n}$ and normalized by dividing by $\sqrt{n}$. As $n
\rightarrow \infty$, $z$ converges in distribution to the uniform
distribution on the disk $|z|<1$.  Furthermore, as $n \rightarrow
\infty$, each eigenvalue is almost surely non-real.
\end{theorem}


\section{Random polynomials and companion matrices}

Let us recall again why we are concerned with the special
case of random polynomials; the linear part of the derivative of
time-delay dynamical systems --- the ones often used to fit real
time-series data --- are companion matrices.  In particular, given the
companion matrix:  
\[ \left[ \begin{array}{ccccccc}
      a_{1} & a_2 & a_3 & \cdots & a_{d-2} & a_{d-1} & a_{d} \\
      1 & 0 & 0         & \cdots & 0 & 0 & 0 \\
      0 & 1 & 0 & \cdots         & 0 & 0 & 0 \\
      \vdots & \ddots & \vdots \\
      0 & 0 &0 &\cdots & 0 & 1 & 0 \\

\end{array} 
\right]\] 
the corresponding characteristic polynomial --- the equation whose
solutions are the eigenvalues of the above matrix --- is a polynomial as
given in equation \ref{eqn:standard_poly}.  Thus, the elements of a
companion matrix (e.g. the $a_k$'s) can be identified
with the coefficients of the characteristic polynomial of the given
matrix.  The results we will
discuss here briefly are from \ref{sec:gaussian_poly_thry} while
issues that arise due to computation of eigenvalues from companion
matrices can be found in \cite{compute_poly_zeros}.  


\subsection{Polynomials with Gaussian coefficients}
\label{sec:gaussian_poly_thry}

Let us begin with the polynomial
\begin{equation}
\label{eqn:standard_poly}
a_0 + a_1 x + a_2 x^2 + \cdots + a_n x^n
\end{equation}
where the $a_i$ coefficients are independent standard normals with mean
zero.  The expected number of real zeros, $E_{real}$, as $n
\rightarrow \infty$ is given by the formula:
\begin{equation}
\label{eqn:exp_poly_real}
E_{real}(n) = \frac{2}{\pi} \log (n) + C_1 + \frac{2}{n \pi} + O(1/n^2)
\end{equation} 
where $C_1 = 0.6257358072$ (cf. theorem $2.1$ in
\cite{edeman_poly_zeros} or \cite{kac_poly1}).  This formula is
calculated by integrating the true density which is given by:
\begin{equation}
\label{eqn:comp_real_density}
\rho_d (x) = \frac{1}{\pi} \sqrt{\frac{1}{(x^2 -1)^2} - \frac{(d+1)^2
    x^{2d}}{(x^{2d+2} -1)^2}}
\end{equation}
Careful analysis of equation \ref{eqn:comp_real_density} yields the
limiting density of real eigenvalues --- that as $n \rightarrow
\infty$, the density of real eigenvalues is concentrated at $\pm 1$.
The expected value of real roots is, of course, a crude measurement of
interest as we are interested in differentiating between bifurcations
due to real and complex eigenvalues.  The density provides
considerably more insight.  Moreover, the convergence of the density
will have a significant impact on the probability of the first
bifurcation as we will see in section \ref{sec:numerical_cases} where
we compare equation \ref{eqn:comp_real_density} with the numerical
results.


\subsubsection{Non-zero mean}

\begin{figure}
\begin{center}
\epsfig{file=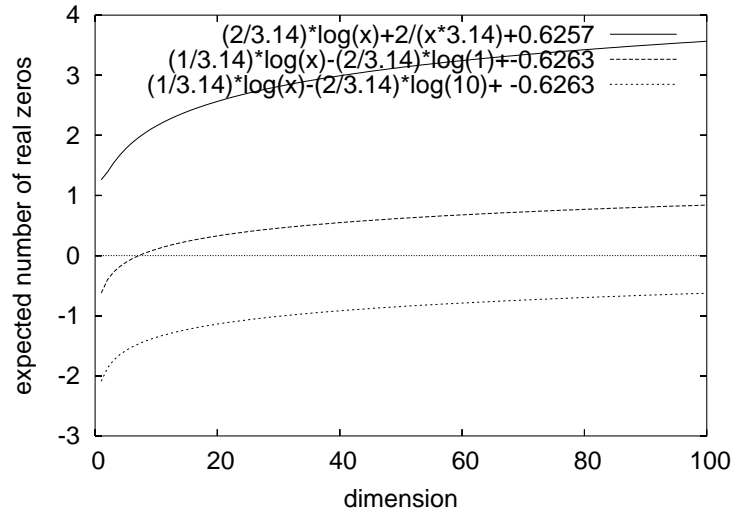, height=10cm, angle=270}
\end{center}
\caption{A plot of equation \ref{eqn:exp_poly_real} and \ref{eqn:exp_poly_real_nonzero} for $m=1,
10$.}
\label{fig:companion_poly_nonzeromean}
\end{figure}

The random polynomials is the only case we will consider for which
there are theoretical results regarding the convergence of the density
of real zeros for distributions with non-zero means.  The effect of a
non-zero mean on the distribution of the coefficients of a random
polynomial is neatly summarized by the following theorem:
\begin{theorem}
Consider a random polynomial of degree $d$ with coefficients that are
independent and identically distributed normal random variables.
Define $m\neq 0$ to be the mean divided by the standard deviation.
Then, as $d \rightarrow \infty$,
\begin{equation}
\label{eqn:exp_poly_real_nonzero}
E_{real}(d) = \frac{1}{\pi} \log(d) + \frac{C_1}{2} + \frac{1}{2} -
\frac{\gamma}{\pi} - \frac{2}{\pi} \log(|m|) + O(1/d)
\end{equation}
where $C_1 = 0.6257358072...$ as previously defined, and $\gamma =
0.5772156649...$ is Euler's constant.  Furthermore, the expected number
of positive zeros is asymptotic to 
\begin{equation}
\frac{1}{2} - \frac{1}{2} \text{erf }^2 (\frac{|m|}{\sqrt{2}}) +
\frac{1}{\pi} \Gamma [0, m^2]
\end{equation}
\end{theorem}
The proof can be found in \cite{edeman_poly_zeros} (theorem $5.3$).
Considering equation \ref{eqn:exp_poly_real_nonzero} and comparing it
with equation \ref{eqn:exp_poly_real}, one arrives at the comparison
between $\sim \log(d) + \frac{1}{d}$ for zero mean with $\sim \log(d)
- \log(|m|)$; as $d \rightarrow \infty$, the effect of the nonzero
mean is a shift of the $\log(d)$ curve by $-\log(|m|)$.  This
dependence can be seen in Fig. \ref{fig:companion_poly_nonzeromean}.

\subsection{Real zeros of random functions with random coefficients}

 
The goal of this paper is to consider the probability of the type
of bifurcation from a fixed point in a class of neural networks that
can be used to approximate time-series data from actual physical
experiments.  In \cite{edeman_poly_zeros} a remarkable theorem was
proved regarding sums of differentiable functions with random
coefficients linking the distribution to the coefficients and the
function type to the distribution of real zeros.  This theorem has a
high degree of relevance to a study such as ours since it offers the hope of
applying to the class of mappings we are considering.
\begin{theorem}
\label{theorem:ed_big_gun}
Let $v(x) = (f_0(x), \dots, f_n(x))^T$ be any collection of
differentiable functions and $a_0, \dots, a_n$ be the elements of a
multivariate normal distribution with mean zero and covariance matrix
$C$.  The expected number of real zeros on an interval (or measurable
set) $I$ of the equation
\begin{equation}
a_0 f_0 (x) + a_1 f_1(x) + \dots + a_n f_n(x) =0
\end{equation}
is
\begin{equation}
\int_I \frac{1}{\pi} ||w'(x)|| dx,
\end{equation}
where $w$ is given by
\begin{equation}
w(x) = \frac{C^{1/2} v(x)}{||C^{1/2} v(x) ||}
\end{equation}
In logarithmic derivative notation this is
\begin{equation}
\frac{1}{\pi} \int_I (\frac{\partial^2}{\partial x \partial y} (\log
(v(x)^T C v(y))|_{y=x=t}))^{1/2} dt
\end{equation}
\end{theorem}
There are many applications of this profound theorem presented in
\cite{edeman_poly_zeros}.  One of particular interest is an application
to a trigonometric series such as
\begin{equation}
\sum_{k=0}^{\infty} a_k \cos (\nu_k \theta) + b_k \sin (\nu_k \theta)
\end{equation}
where $a_k$ and $b_k$ are independent normal random variables with mean
zero and variance $\sigma_k^2$.  The density and, thus, the expected values
of real zeros is quite easily computed given theorem
\ref{theorem:ed_big_gun}.  The density is constant and, therefore, the real
zeros  of the above random trigonometric sum are uniformly distributed
on the real line.  An important point to note is the sharp difference
between the trigonometric case and the random polynomial case.  Thus,
different functional forms and distributions of $a_k$'s give very
different distributions of real zeros of polynomials.  These two
results show a sharp contrast regarding bifurcations from fixed
points --- the trigonometric series above is likely to yield all real
bifurcations; whereas, the standard random polynomial case yields
significantly more bifurcations due to complex eigenvalues.  Providing a
similar analysis for neural networks like those given in equation
\ref{equation:intro_net} is complicated by the sum inside the
activation function and, therefore, is beyond the scope of this paper.
However, as we will discuss later, because neural networks can
approximate nearly any $C^r$ mapping which include  both the random
polynomials and the trigonometric series, altering the distributions of the
$\beta_i$'s and the $w_{ij}$ will clearly yield very different
densities of real zeros.  A relation between the distribution of the $\beta_i$'s and
$w_{ij}$'s will be the focus of section \ref{sec:neural_net_numerical}.



\section{A conjecture regarding the first bifurcation from a fixed point}
\label{sec:firstbifurcationconjecture}
There are three generic, codimension-one, local bifurcations from a
fixed point in maps of dimension two or greater \cite{kuzbook} \cite{brunovsky_generic}.  These
three bifurcations depend on symmetries of the dynamical system, but
generally they consist of: the flip
bifurcation, corresponding to the largest eigenvalue being $-1$; the
fold, corresponding to the largest eigenvalue being $1$; and the
Neimark-Sacker \cite{naimark} \cite{sacker}, corresponding to a
complex conjugate pair of eigenvalues with modulus one.  Edelman, Girko, and Bai have all
shown that in the infinite-dimensional limit, a real matrix with
elements selected from a real Gaussian distribution, the normalized
eigenvalues will be distributed uniformly on the unit disk in the
complex plane.  Since the Neimark-Sacker bifurcation corresponds to
the bifurcation via a complex conjugate pair of eigenvalues, a logical
application of the circular law is to infinite-dimensional dynamical
systems whose Jacobian matrix has elements whose distribution has a
finite sixth moment.  In this circumstance, the probability one
bifurcation would seem to be a Neimark-Sacker bifurcation.  A conclusion along these lines will prove
incorrect as we will show in section \ref{sec:numerical_cases}.  The
restriction of a finite sixth moment will turn out to be too weak
because lower-order moments can affect quantities like the spectral
radius or the eigenvalue with the largest modulus.  Limiting ourselves
to the case where the matrix has real Gaussian elements with mean zero
and unit variance  for which we have more restricted and detailed
results will be fruitful.  For example, if the real
eigenvalues concentrate near $1$ and $-1$, we will run into problems, but Edelman (corollary (\ref{corollary:fourpointfive})) has shown
that this circumstance will not occur.  Instead, the
real eigenvalues will be distributed uniformly on the real axis.
Results akin to the aforementioned result will prove necessary for
arguments involving bifurcations.  In
the standard and general bifurcation sequence constructions, one would be
concerned with a parameterized curve of matrices.  In such a scenario
the matrices would not be independent along the curve in general.
Surmounting this obstacle is yet an open problem.  However, in some
special cases, like where the parameterized curve is
linear and forms an interval in say, $R^1$, the difficulties are greatly reduced.  Thus, we can make the following statement:
\begin{corollary}{ \bf ( First bifurcation probability ) }
\label{corollary:firstbifurcation}
Given the dynamical system $F$ 
\begin{equation}
F(x_{t-1}) = x_t = \epsilon A x_{t-1} + \epsilon G(x_{t-1})
\end{equation}
where $x_t \in R^n$, $\epsilon \in R$, $A \in R^{n^2}$, $a_{ij}
\in N(0, 1)$, and where $G(x_{t-1})$ is a nonlinear $C^r$ ($r>0$)
mapping of $x_{t-1}$ which is of order $2$ or higher.  Thus
$F(x_{t-1}) \cong \epsilon A x_{t-1}$ for $\epsilon$ small.  Assume
$F$ has a fixed point at $\epsilon = 0$ and upon the increase of $\epsilon$, $F$ undergoes a local,
codimension-one bifurcation.  As the dimension of the dynamical system
$F$ goes to infinity (i.e. given $A \in R^{n^2}$, $n \rightarrow
\infty$), the probability that the first bifurcation will be of type
Neimark-Sacker will converge to one.
\end{corollary}   
\textit{Proof:} This result follows trivially from the results of
Edelman \cite{edelman1}, \cite{edelman2} and Girko \cite{girko1}
\cite{girko2} \cite{girko3} \cite{girko4} \cite{girkobook}.
 
We can, with a little work, impose a measure on the set of dynamical
systems for which this result holds via results of Edelman, the neural
networks\footnote{In theory we can impose a measure on the weights of the neural networks
  that mimics the results of random matrices, however we have not
  found this said measure.  It is likely that it could be determined
  via training of neural networks on data from dynamical systems with
  random ($N(0, 1)$) linear parts.}, and some standard arguments using measure theory.  Upon
doing so, one nontrivial issue is understanding what such a set of
dynamical system would  ``look'' like.  We will
refrain from further discussion of this extension here.  It was
originally hoped that we could extend
this result such that the elements of the $A$ matrix can be selected from
any distribution with a finite sixth moment in line with the circular
law of Bai \cite{bai}.  However, considering only random matrices with
elements chosen from a Gaussian distribution, it can be shown
numerically that the mean of the distribution has a significant effect
on the eigenvalue with the largest modulus.

Corollary (\ref{corollary:firstbifurcation}) falls far short of
satisfying desires.
First, corollary (\ref{corollary:firstbifurcation}) does not speak to
the probability of its
hypothesis being satisfied in a general space of dynamical
systems.  Again, we can construct a measure such that the relevant
hypothesis are always satisfied, but how general such a measure is,
which is the same problem, is still a problematic issue.  Moreover,
corollary (\ref{corollary:firstbifurcation}) is not cast in the
general parameterized curves of the bifurcation theory framework we
desire --- linear ``curves'' are very limiting.  See
\cite{soyomayorbifsets} or \cite{globalwiggins} for a construction of
the general bifurcation framework to which we are referring.
Providing an answer to the question regarding general spaces of $C^r$
mappings would likely require
establishing some type of measure on the space of general $C^r$
dynamical systems, and establishing such a measure is
often very difficult, to say the least.  If such a measure could be defined,
some meaningful notion of equivalence in measure would need to be
shown.  This issue, however,
would likely be the least of the problems.  Moreover, corollary (\ref{corollary:firstbifurcation}) does not provide any information
regarding how large, but finite-dimensional, dynamical systems that
satisfy all the hypotheses behave.  Lastly, corollary
(\ref{corollary:firstbifurcation}) does not provide any insight into
how the convergence to such a result might occur as the dimension of
the dynamical system is increased.  Thus, we will present a conjecture
that we believe captures more of what we want.
\begin{conjecture}{ \bf (Genericity of Neimark-Sacker bifurcations in high-dimensional dynamical systems) }
\label{conjecture:firstbifurcation}
Begin with a space of $C^r$ $(r>0)$ dynamical systems with bounded
first derivatives whose elements form a distribution with zero mean on compact sets with a single real parameter and such that there
exists a fixed point on a measurable interval of the parameter
space, and at least one local
bifurcation upon a continuous variation of the parameter.  There
exists a probability measure on the parameters such that, as the
dimension $d$ of the
dynamical system is increased, the probability of the bifurcation from
fixed points via the Neimark-Sacker bifurcation will increase and
approach probability one as $d \rightarrow \infty$.
\end{conjecture}
Of course this conjecture makes the most sense in a framework such as that
provided by neural networks for which the parameter space is large
enough that it can have some approximation of $C^r$ function space.
We will not discuss this conjecture further here, but upon the
presentation of the numerical results, we will provide a discussion of
what this conjecture might mean and where and how it is known to fail.

Finally, the above construction does not apply explicitly to the time-delay
dynamics in general --- whose linear derivative matrices are companion
matrices --- or to the neural networks we are focusing on in particular.
Such a circumstance would require a slightly different conjecture:
\begin{conjecture}{ \bf (Genericity of Neimark-Sacker bifurcations in
    high-dimensional time-delay dynamical systems) }
\label{conjecture:firstbifurcation_comp}
Begin with the space of $C^r$ $(r>0)$ time-delay dynamical systems with bounded first derivatives on compact sets with a single, real parameter and such that there
exists a fixed point on a measurable interval of the parameter
space, and at least one local
bifurcation upon a continuous variation of the parameter.  There
exists a probability measure on the parameters such that, as the
dimension $d$ of the
dynamical system is increased, the probability of the bifurcation from
fixed points via the Neimark-Sacker bifurcation will increase, and
approach probability one as $d \rightarrow \infty$.
\end{conjecture}
Analysis of this conjecture is akin to a study of random polynomials or a sum of
random functions with coefficients drawn from a random distribution.
This case is of interest because many experimental results come from
time-series data --- data that is reconstructed with a ``universal
approximator'' that forms a time-delay dynamical system (neural
networks of the type we consider fall into this class).  We will
compare and contrast the various results and implications of the constructions of the two conjectures.  Because
neural networks are universal approximators, we will, at the end,
suggest how to find the measure on the $a_k$'s that will link them to
the general random matrix framework.


\section{Numerical cases}
\label{sec:numerical_cases}

We will begin our numerical investigations with a careful analysis of
Gaussian random matrices with mean zero and variance $1$ --- one of
the cases for which Edelman has provided theoretical results.  We will
then begin to investigate the difference between, and hence
generality of, what we can imply
regarding conjecture \ref{conjecture:firstbifurcation} with the results of Edelman versus  those of
Bai by perturbing the first and second moments of the Gaussian
matrices and observe differences.  Following this, we will study the
case of uniform random matrices --- a case for which Bai's results
still apply --- to begin to understand the invariance of various
results to different random distributions.  We will again study the
perturbation of the first and second moments of the uniform
distribution on the relevant quantities.  Since the Jacobian of a
time-delay dynamical system at a fixed point is a companion matrix, we
will then briefly study companion matrices and finally move on to
the case of time-delay neural networks.





\subsection{$M_n$ Gaussian}
\label{sec:m_gaussian}

As a base case, we will begin our analysis with matrices $G$ that have
elements chosen from a Gaussian distribution with mean zero and
variance one ($N(0, 1)$).  This is the case that both Girko and Edelman
investigated, and thus there exist several theorems upon which we can
verify our work.  We will then begin perturbing various moments of this
distribution to better understand the different implications and
limitations of the work of Edelman, Girko, and Bai as applied to
bifurcation theory. 


\subsubsection{Gaussian matrices with zero mean}

\begin{figure}[tbp]
\begin{center}
\epsfig{file=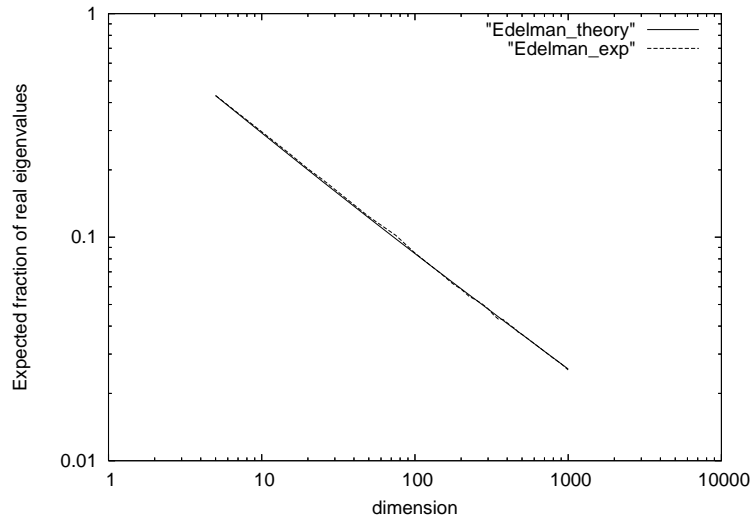, height=10cm, angle=270}
\end{center}
\caption{Edelman's prediction for the expected fraction of real
  eigenvalues and the empirically calculated expected number of real
  eigenvalues.  Both quantities were calculated in increments of $5$
  dimensions up to $d=50$ and then increments of $25$ thereafter until
  $d=1000$.  The line is that of a power law with $E_{\lambda_{real}}[d]
  \sim 0.9784 d^{-0.5291}$ as expected.}
\label{fig:edelman_gaussian_predictions}
\end{figure}

Begin with the mean zero, variance one case, a comparison between Edelman's expected number of real eigenvalues given
in equation \ref{equation:ed_formula} and those empirically calculated
from random matrices is depicted in
Fig. \ref{fig:edelman_gaussian_predictions}.  The lines overlay nearly
perfectly even at low dimensions yielding a power-law dependence of
$E[\lambda_{real}] = 0.978 d^{-0.529}$ as expected.    

\begin{figure}[tbp]
\begin{center}
\epsfig{file=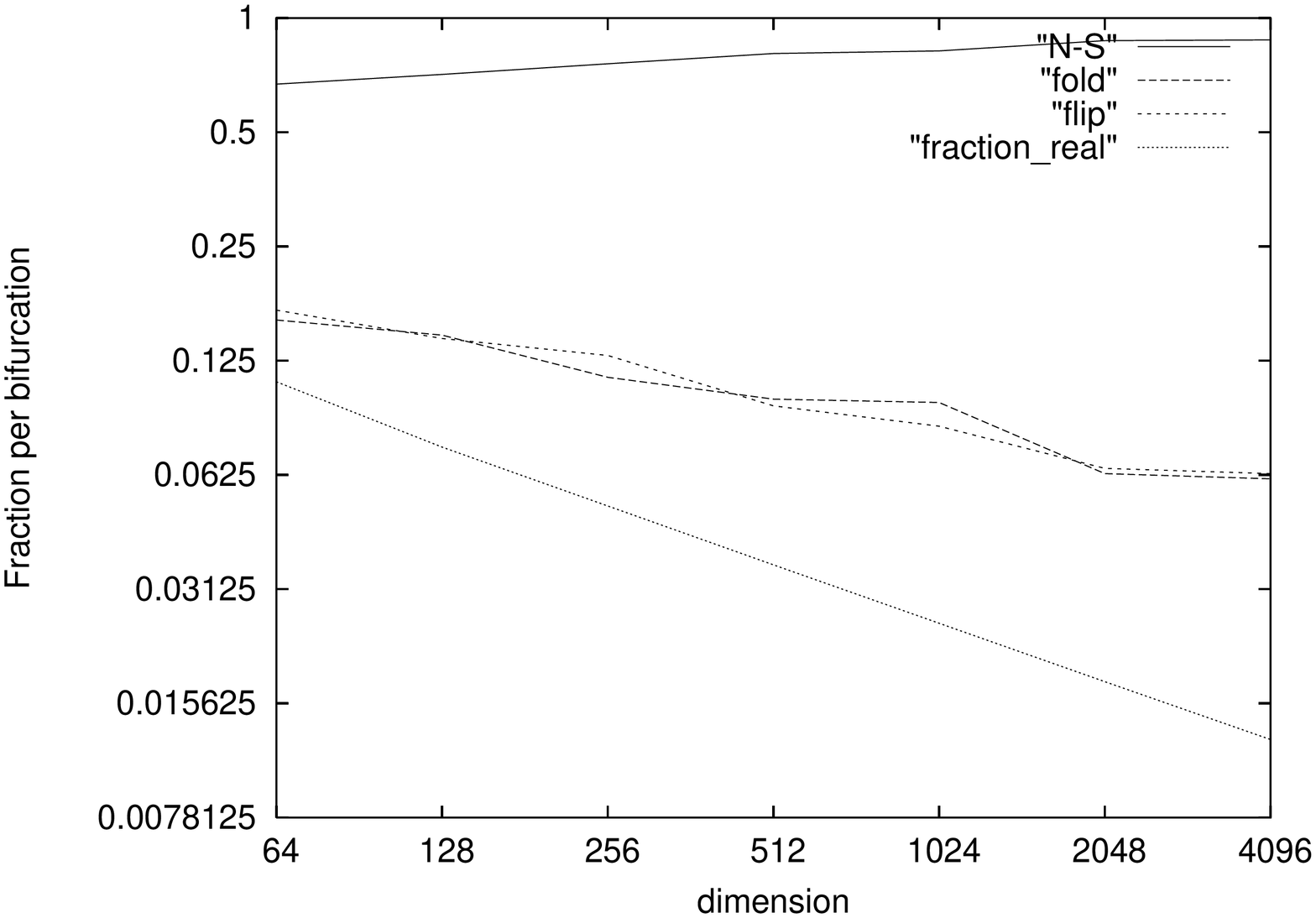, height=8.5cm, angle=270}
\epsfig{file=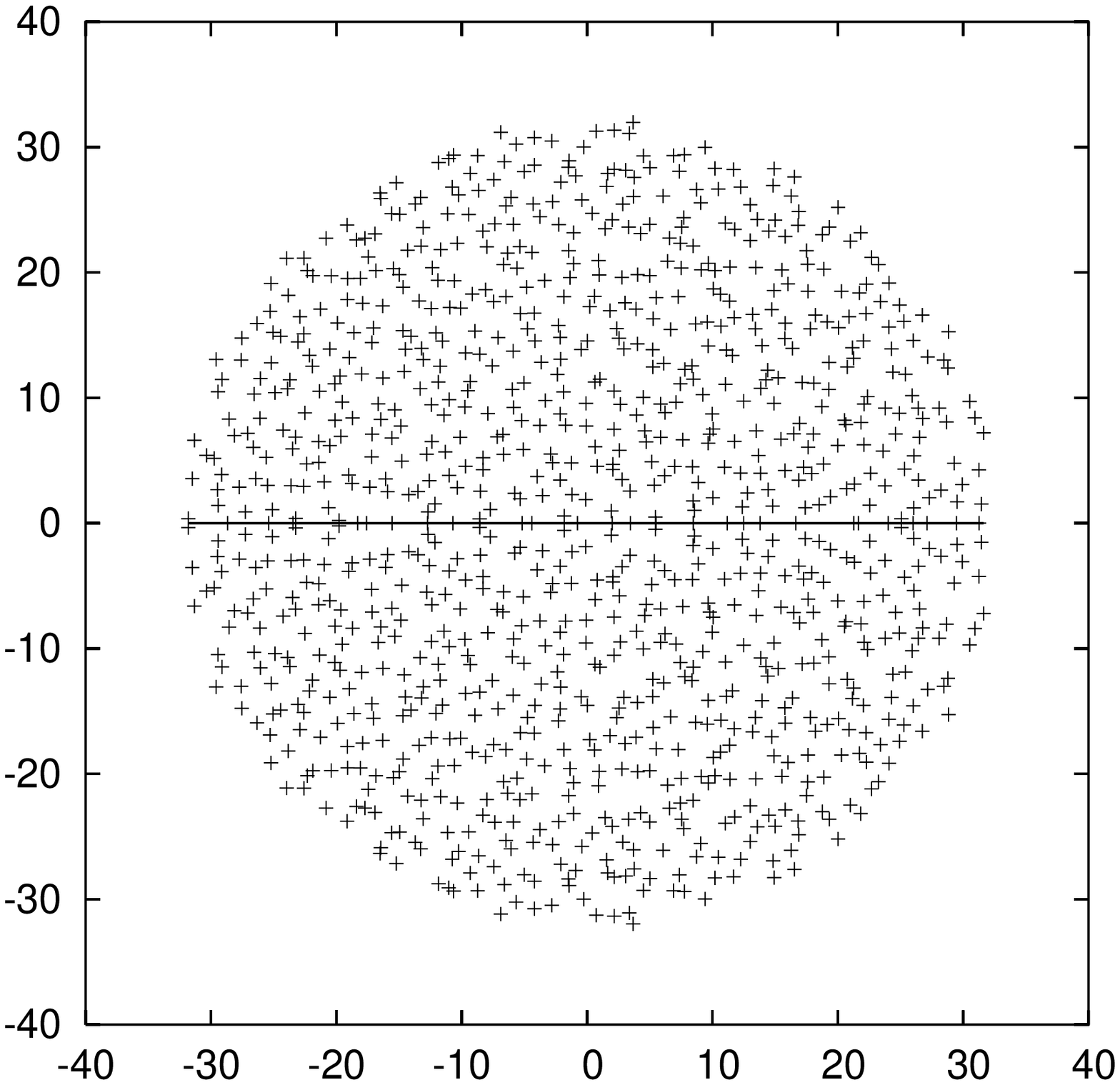, height=8.5cm, angle=270}
\end{center}
\caption{On the left, the observed probability of each bifurcation was
recorded for $1000$ matrices with i.i.d., mean zero, variance one, Gaussian elements for each $d$ (in powers of $2$) along
with the fraction of eigenvalues that are real.  On the right is the
spectrum of eigenvalues in the complex plane that corresponds to a
single $1024 \times 1024$ matrix ($d=1024$).}
\label{fig:gaussian_bif_fraction}
\end{figure}

Given corollary \ref{corollary:fourpointfive} --- that the
distribution of real eigenvalues converges to a uniform distribution
on the real line --- a disproportionate fraction of the bifurcations
should be due to complex eigenvalues especially as $d \rightarrow
\infty$.  Figure \ref{fig:gaussian_bif_fraction} depicts the fraction
of bifurcations that correspond to Naimark-Sacker, flip, and fold
bifurcations as well as the fraction of eigenvalues that are real.  For systems we are considering, these distinctions are
easily made by simply calculating the largest eigenvalue and
determining whether it is complex or real, combined with the sign of the
eigenvalue.  The results are as one might expect in the sense that the
number of real bifurcations due to positive and negative eigenvalues
are nearly identical.  However the fraction of real eigenvalues
converges to zero considerably faster than the fraction of
bifurcations due to real eigenvalues.  Considering the right plot of
Fig. \ref{fig:gaussian_bif_fraction}, the spectrum of eigenvalues of a
$1024 \times 1024$ matrix, the real line is clearly highly and evenly
populated with $\sim 30$ real eigenvalues.  Hence the convergence to
uniformity seems to be well-behaved by $d=1024$.  Nevertheless, the fraction of
bifurcations due to real eigenvalues is clearly decreasing with $d$ in
a  power law.  Moreover, the decrease of bifurcations due to $1$ and
$-1$ are nearly identical as is expected.

\subsubsection{Perturbing the first moment --- Gaussian matrices with
  a non-zero mean}
\begin{figure}[tbp]
\begin{center}
\epsfig{file=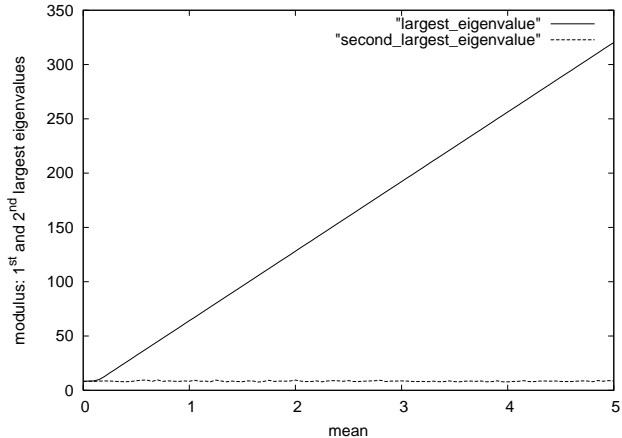, height=8.5cm, angle=270}
\end{center}
\caption{This figure represents an ensemble of $1000$ $d \times d$
  matrices with $d=64$.  Depicted are the modulus of largest and second largest
  eigenvalues.  The line representing the modulus of the largest
  eigenvalue is given by $64 m$ while the line for the modulus of
  the second largest eigenvalue is given by $\sim \sqrt{d}$.}
\label{fig:gaussian_nonzero_mean}
\end{figure}

The results of Girko and Edelman pertain to real random matrices with
Gaussian matrices with mean zero and finite variance, while the results
of Bai apply to any real random matrix with a distribution with a
finite sixth moment.  There is a great difference between these two formulations.  Beginning with a standard Gaussian
random matrix and perturbing the first moment (the mean), yields a
result that has little relevance for the random matrix theory but has
significant implications from the prospective of bifurcation theory.
Summing a $d \times d$ matrix $G$ as a matrix with elements $g_{ij}$ drawn from a
Gaussian distribution with mean zero and variance one ($N(0, 1)$) and
a constant $d \times d$ matrix $P_m$ with elements $p_{ij}=m$ yielding 
\begin{equation}
\Lambda_m = G + P_m
\end{equation}
Figure \ref{fig:gaussian_nonzero_mean} has the modulus of the largest and
second largest eigenvalues plotted for $\Lambda_m$ with $d=64$.  For
$|m| > 0.1289$, the largest eigenvalue is always real and increases with
$m$ according to:
\begin{equation}
\lambda_d (m) = d m
\end{equation}
The next largest eigenvalue is most often complex as would be expected
if the distribution of eigenvalues on the unit disk is uniform.
However, the modulus of the second largest eigenvalue(s) is constant independent of $m$ ($\lambda_{d-1} = \text{constant}$) and is
approximately $\sqrt{d}$ as expected from considering Edelman's
normalization formulas.  Thus, aside from the largest real eigenvalue, the distribution of
eigenvalues behaves like the spectrum of $G$.  Therefore, a dynamical
system with a $DF$ matrix
\begin{equation}
\frac{\epsilon}{\sqrt{d} + a} \Lambda_{m=\frac{1}{d}}
\end{equation}
for $a<<1$, upon increasing $\epsilon$, will always undergo a flip
bifurcation from a fixed point.  The point of this is that Bai's
convergence of the spectrum to a uniform distribution on the unit disk
in the complex plane of a random matrix with elements drawn from a
distribution with finite sixth moment is not sufficient to guarantee that
a dynamic system with a Jacobian matrix (at a fixed point) with
elements that converge to a distribution with a finite sixth moment,
will bifurcate as a matrix that was defined by having eigenvalues in a
uniform distribution on the unit disk for all $d$.  In this case, a measure-zero set (a single eigenvalue), happens to
be of utmost importance if considering the most probable bifurcation
from a fixed point.  Moreover, this single eigenvalue is not a counter-example to Bai's theorem because aside
from the one single eigenvalue, the rest of the spectrum converges
uniformly on the unit disk in the complex plane.  


\subsubsection{Perturbing higher moments}

Perturbation of the second moment of the distribution of $G$, which
amounts to perturbing the variance, has a simple effect on the
spectrum --- the spectral radius increases linearly with the
variance.  This will have very little effect on the bifurcation
structure aside from decreasing the $\epsilon$ value for which the
first bifurcation from a fixed point will occur.  Perturbing higher
moments of the distribution and the subsequent effects remains an open problem.

\subsection{$M_n$ uniform}
\label{sec:m_uniform}


Above we considered matrices with entries drawn from Gaussian
distributions.  As a result, we could bring to bear a large body of
analytical machinery.  Now, however, we turn to matrices with entries
drawn from iid uniform random variables on the interval $(a,b)$.
Unfortunately, in this circumstance, very little analytical machinery
is available.

\subsubsection{Uniform matrices with zero mean}
\begin{figure}[tbp]
\begin{center}
\epsfig{file=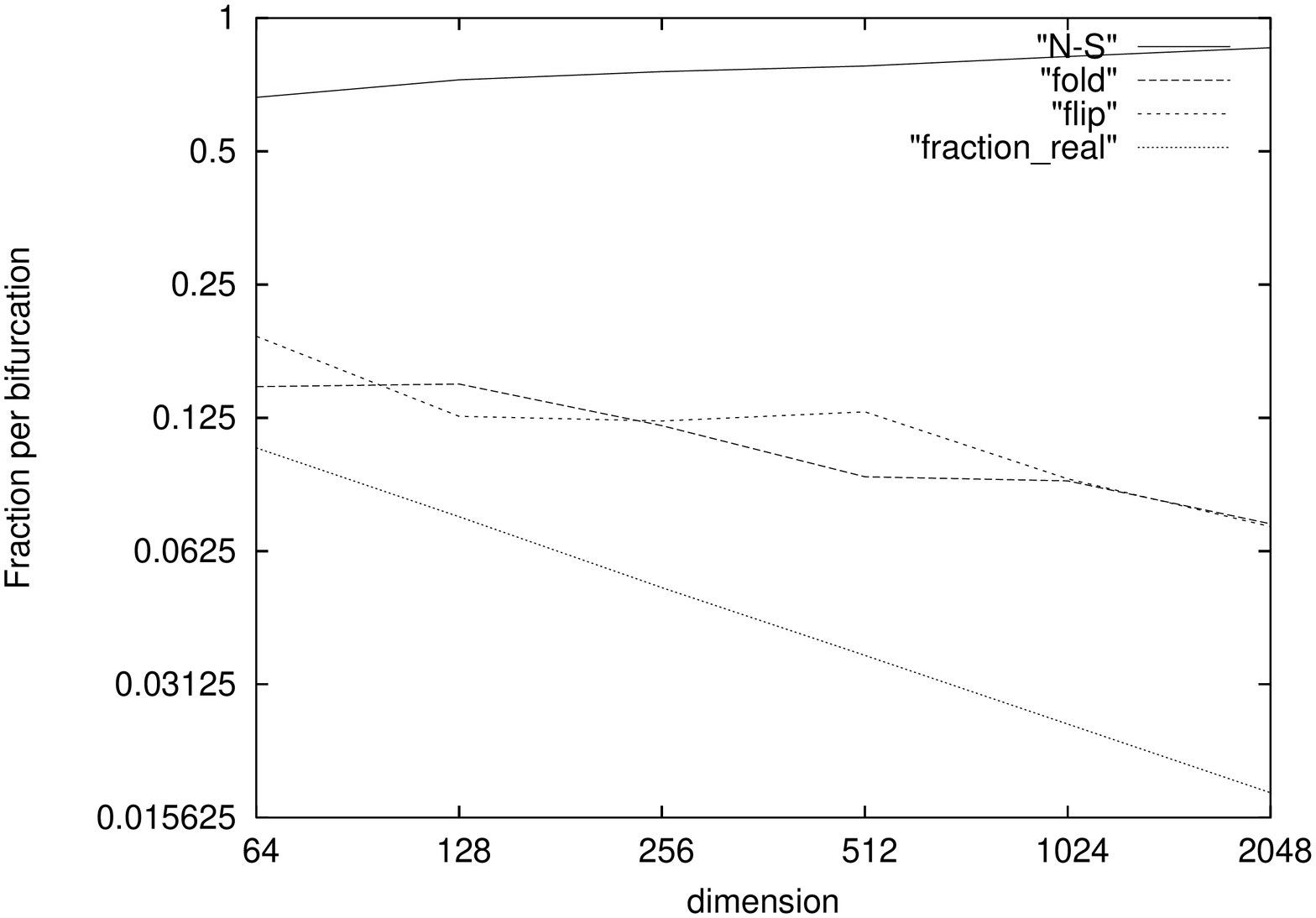, height=8.5cm, angle=270}
\epsfig{file=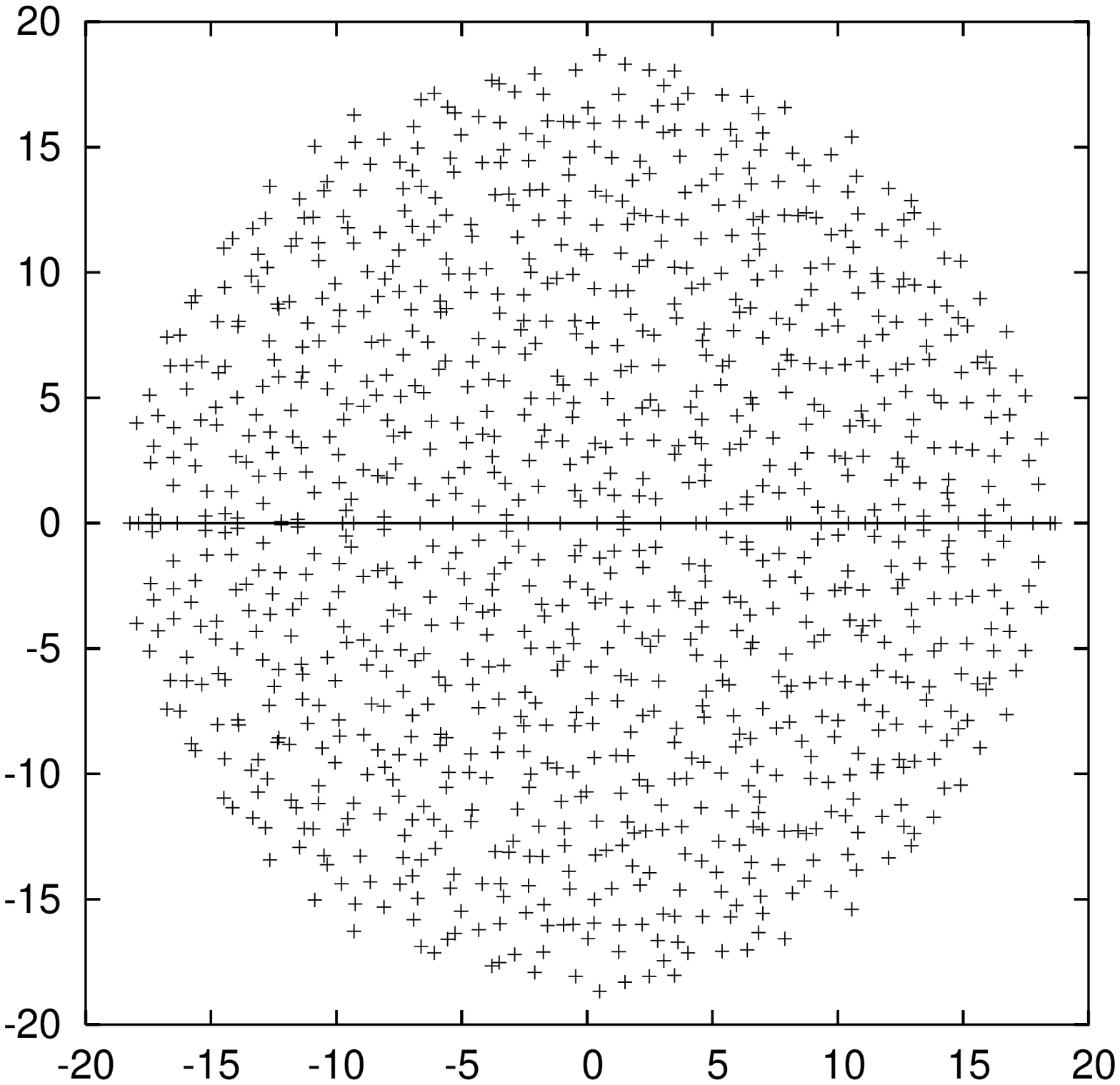, height=8.5cm, angle=270}
\end{center}
\caption{On the left, the observed probability of each bifurcation was
recorded for $1000$ matrices  with i.i.d. uniform elements for each $d$ (in powers of $2$) along
with the fraction of eigenvalues that are real.  On the right is the
spectrum of eigenvalues in the complex plane that corresponds to a
single $1024 \times 1024$ matrix ($d=1024$).}
\label{fig:uniform_bif_fraction}
\end{figure}

For simplicity, we will set $a = -1$ and $b=1$ such that the
$u_{ij}$'s are i.i.d. uniform random variables on $(-1, 1)$, with mean
zero.  Concentrating first on the fraction of real eigenvalues,
Fig. \ref{fig:uniform_bif_fraction} yields a scaling law of
$E[\lambda_{real}] = 0.919 d^{-0.517}$.  This is similar to the
theoretical and empirically calculated $E[\lambda_{real}]$ for a
random matrix with Gaussian elements with zero mean.  However,
$E[\lambda_{real}]$ approaches zero faster in the case of Gaussian
random matrices than with uniform random matrices.  The difference is,
nevertheless, quite small (about $3$ percent).  Accordingly, the fraction
of real bifurcations decreases slightly slower for uniform random
matrices than for the Gaussian case, while the fraction of flip and
fold bifurcations are identical up to standard error.  The right plot
in Fig.  \ref{fig:uniform_bif_fraction} is indistinguishable from the
analog plot for Gaussian random matrices given in
Fig. \ref{fig:gaussian_bif_fraction}.  The real line is highly and evenly
populated with $\sim 25$ real eigenvalues.

\subsubsection{Perturbing the first moment --- Gaussian matrices with
  a non-zero mean}

\begin{figure}[tbp]
\begin{center}
\epsfig{file=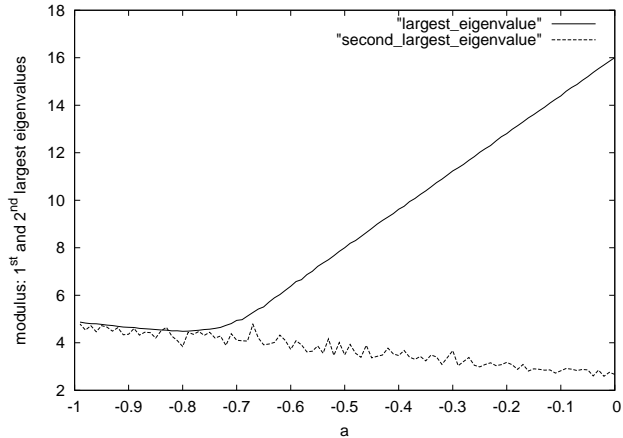, height=8.5cm, angle=270}
\end{center}
\caption{This figure represents an ensemble of $1000$ $d \times d$
  matrices with $d=64$ per $a$ increment where $\delta a = 0.01$.  Depicted are the modulus of largest and second largest
  eigenvalues.  The line representing the modulus of the largest
  eigenvalue is given by $\frac{a}{4}$.}
\label{fig:uniform_nonzero_mean}
\end{figure}

The uniform random variable case is fundamentally different from the
Gaussian case in several ways.  First, the uniform case has finite
support on $R$.  Second, the mean is  directly related to the
endpoints of the support of the distribution, i.e. $\bar{x} =
\frac{b+a}{2}$.  Thus perturbing the mean from zero amounts to
upsetting the symmetry of the end points of the support about zero.  To investigate the effects of perturbing the mean we fix $b=1$ and
vary $a$ from $0$ to $-1$.  Or, more explicitly, given
\begin{equation}
\Lambda_a = U + P_a
\end{equation}
with $u_{ij}$ uniform random variables on $(0, 1)$ and with $p_{ij} =
a$ $\forall i, j$, $a$ is varied on $(-1, 0)$.  Figure \ref{fig:uniform_nonzero_mean}
depicts the dependence of the magnitude of the modulus of the two
largest eigenvalues versus $a$ for a collection of $64 \times 64$
matrices.  For $a=0$ to $a \sim -0.78$ the largest eigenvalue is real
and its magnitude is given by 
\begin{equation}
\lambda_d(a) \sim - \frac{a}{4}
\end{equation}
The knee in the curve which occurs at $\sim -0.78$ for $d=64$ is
dependent upon $b$ --- the upper bound on the support of the
distribution.  Moreover, $\lambda_{d-1}$ does not remain constant
with variation of the mean of the distribution but rather increases.  However, aside from the largest eigenvalue, for $a \in (\sim -0.78,
0)$, the distribution of eigenvalues appears uniform on a disk of
radius $|\lambda_{d-1}|.$  This effect is symmetric about zero.  Fixing $a$ and varying $b$ will
net the same effects.  Thus, for $a \in (\sim -0.78, 0)$, the
corresponding dynamical system will have a $100$ percent probability
of undergoing a fold bifurcation.  This result supports Bai's result because the convergence to uniformity
of the eigenvalues on the unit disc is only violated by a single
eigenvalue which is not relevant for random matrix results regarding
distributions of eigenvalues.  However, it is very relevant from our bifurcation
theory prospective.

\subsubsection{Perturbing higher moments}
The variance of a uniform distribution on an interval $(a, b)$ is
given by $\text{var} = \frac{(b-a)^2}{12}$.  Thus, increasing the
variance by a factor of $c$ while leaving the mean at zero is identical
to multiplying $a$ and $b$ by $\sqrt{c}$, which is in turn identical
to multiplying $u_{ij}$ by $\sqrt{c}$.  Therefore, increases in the
variance of a mean-zero uniform distribution of the elements of a
random matrix $U$ has the effect of increasing the spectral radius by
a factor of $\sqrt{c}$, which is simply a normalization factor and makes
little difference to either the bifurcation prospective or
the random matrix prospective.  Again, perturbations of moments $>2$
is beyond the scope of this paper.

\subsection{$M_n$ companion Gaussian elements}

\begin{figure}[tbp]
\begin{center}
\epsfig{file=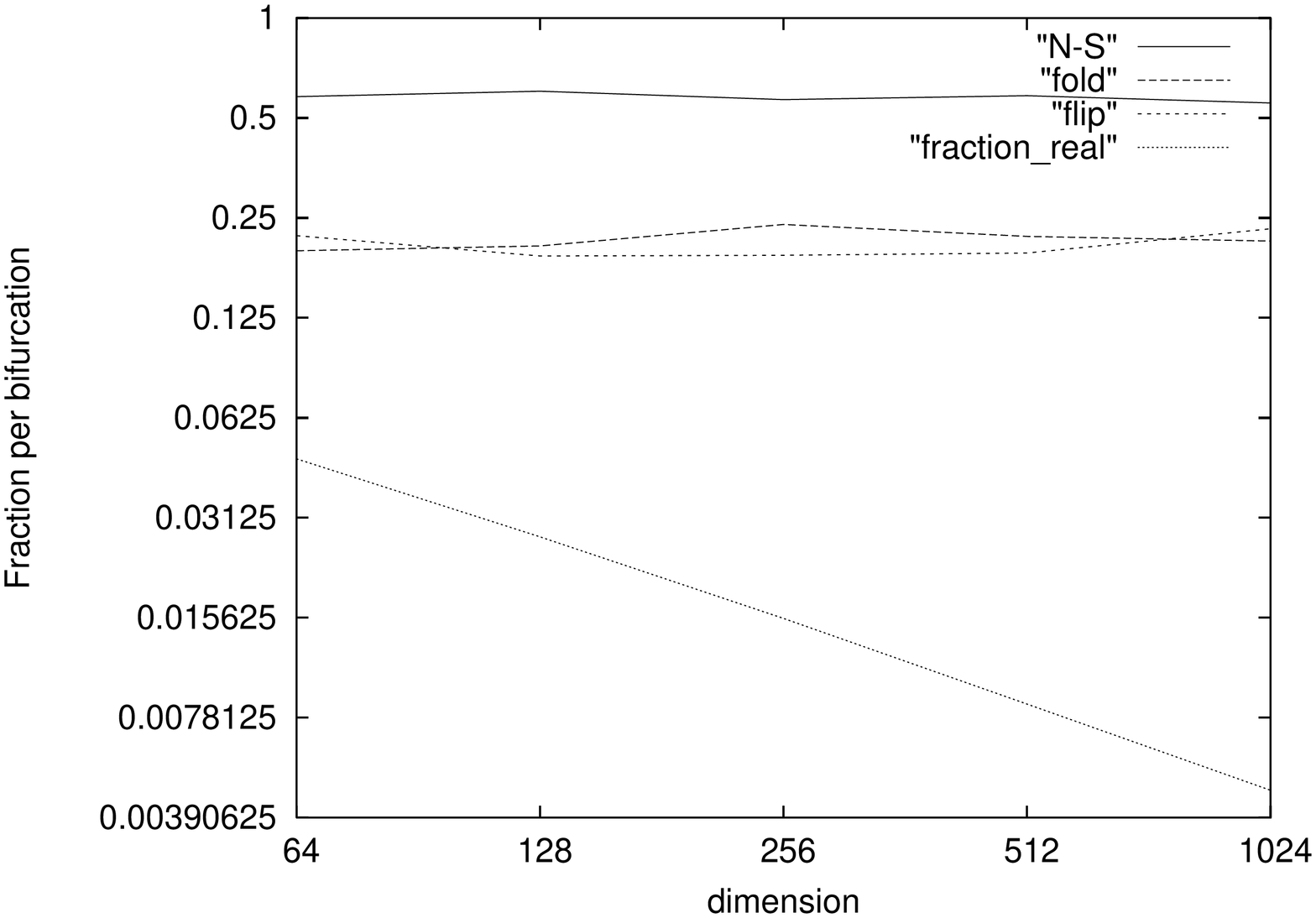, height=8.5cm, angle=270}
\epsfig{file=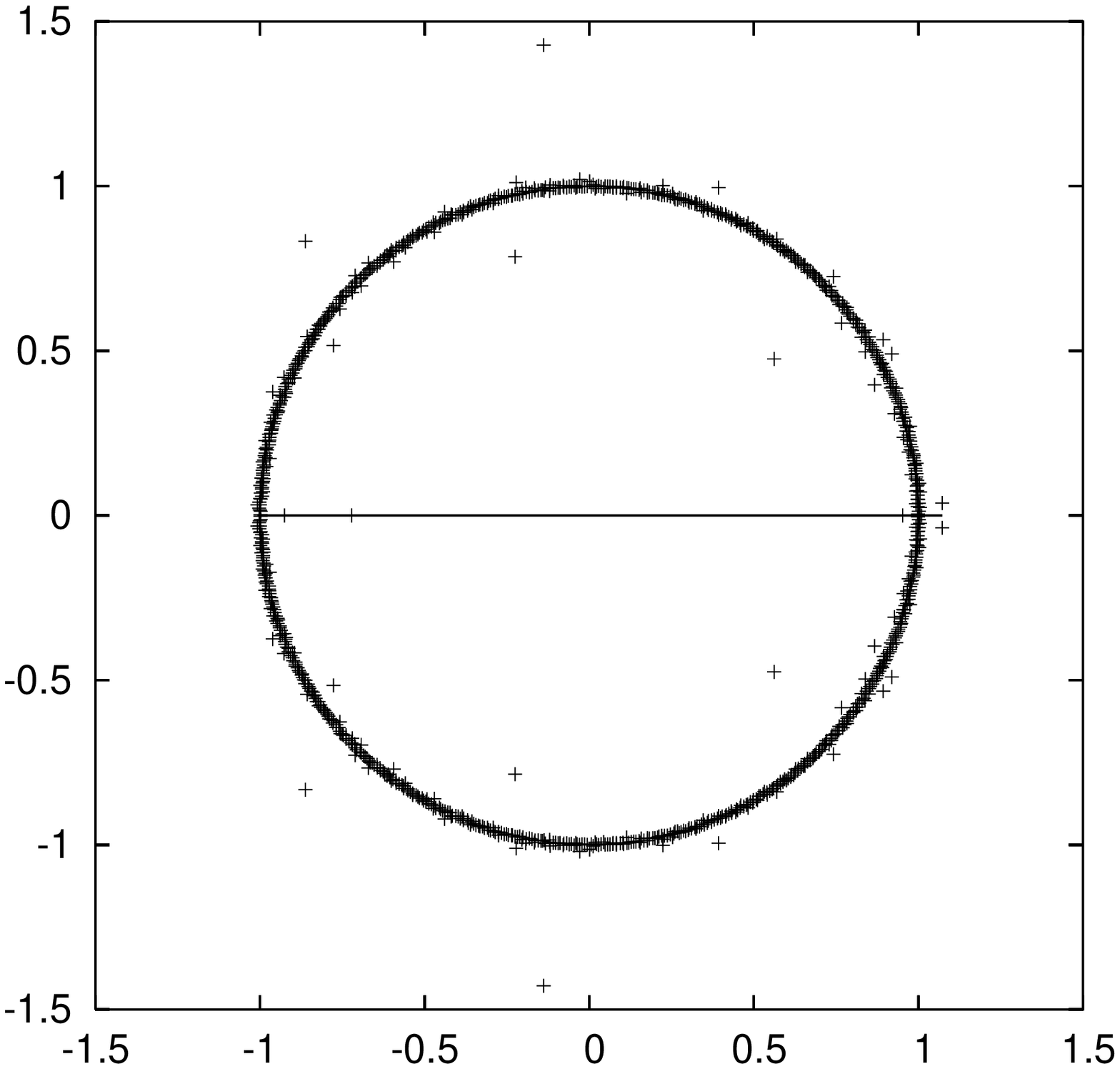, height=8.5cm, angle=270}
\end{center}
\caption{On the left, the observed probability of each bifurcation was
recorded for $1000$ matrices with i.i.d., mean zero, variance one,
Gaussian $a_k$'s for each $d$ (in powers of $2$) along
with the fraction of eigenvalues that are real.  On the right is the
spectrum of eigenvalues in the complex plane that corresponds to a
single $1024 \times 1024$ matrix ($d=1024$).}
\label{fig:companion_bif_fraction}
\end{figure}

Beginning with Fig. \ref{fig:companion_bif_fraction}, there are two
important features.  First, the eigenvalues appear to be distributed
uniformly on the real line and on the unit circle, not on the entire disc.  Second, the
probability of a bifurcation due to a complex eigenvalue hovers near
$55$ percent over a range of $d=128$ to $1024$.  This result, which seems
surprising considering the spectrum depicted in
Fig. \ref{fig:companion_bif_fraction}, can be explained considering the
convergence of the density of real zeros given by Edelman.  This
demonstrates how very important it is to consider the convergence of
the densities when making claims regarding probable bifurcations. 

\begin{figure}[tbp]
\begin{center}
\epsfig{file=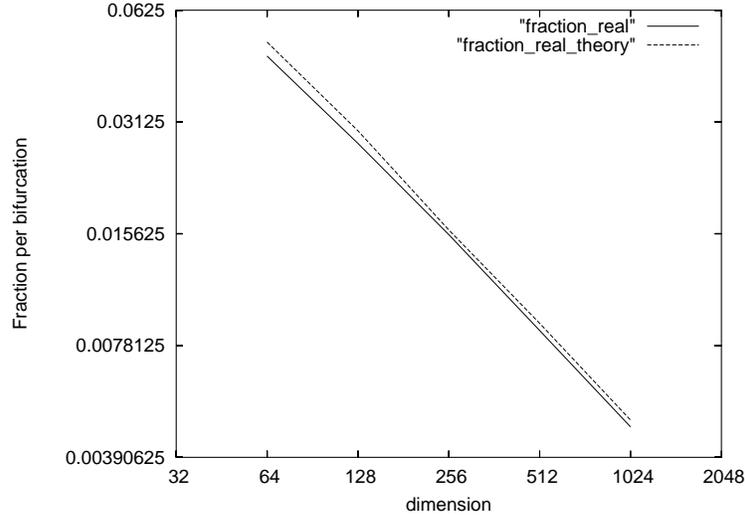, height=10cm, angle=270}
\end{center}
\caption{Edelman's prediction for the expected number of real
  eigenvalues and the empirically calculated expected number of real
  eigenvalues.  Both quantities were calculated for $d=64, 128, 256,
  512, 1024$ and have a clear power-law dependence with
  $\frac{E[\lambda_{real}]}{d} = 1.52 d^{-0.831}$ for the numerically
  generated set and $E[\lambda_{real}]=1.78 d^{-0.849}$ as calculated
  from Edelman's formula (equation \ref{eqn:exp_poly_real}).}
\label{fig:edelman_gaussian_predictions_comp}
\end{figure}

Before we consider the densities, let us begin by considering the
fraction of zeros that are due to real roots.  Figure
\ref{fig:edelman_gaussian_predictions_comp} portrays the numerically
calculated fraction of real eigenvalues for a set of companion
matrices with $a_k$'s drawn from a normal distribution with mean zero
and unit variance along with the predicted fraction of real zeros of
Edelman.  Clearly the two lines are in considerable agreement and have
a power-law dependence with $\frac{E[\lambda_{real}]}{d} = 1.52
d^{-0.831}$ for the numerically generated set and
$E[\lambda_{real}]=1.78 d^{-0.849}$ as calculated from Edelman's
formula given in equation \ref{eqn:exp_poly_real}.  Note that the
falloff of the fraction of zeros that are real is considerably faster
than the other cases considered.  This adds to the surprise found in
Fig. \ref{fig:companion_bif_fraction}.  The fraction of eigenvalues
that are real is decreasing like $d^{-0.85}$, yet the fraction of
bifurcations due to real eigenvalues remains roughly constant for
$d>128$.



\begin{figure}[tbp]
\begin{center}
\epsfig{file=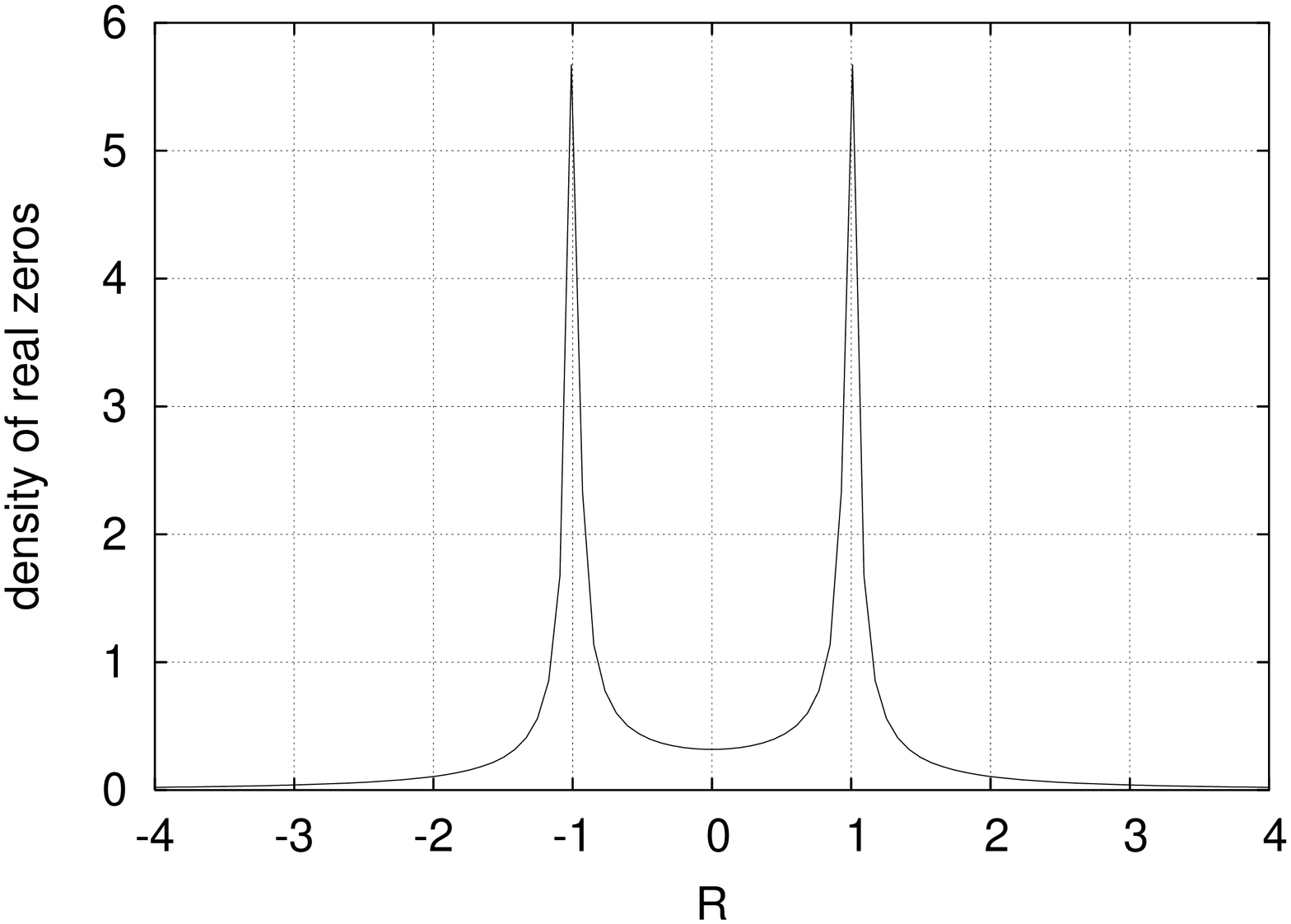, height=8.5cm, angle=270}
\epsfig{file=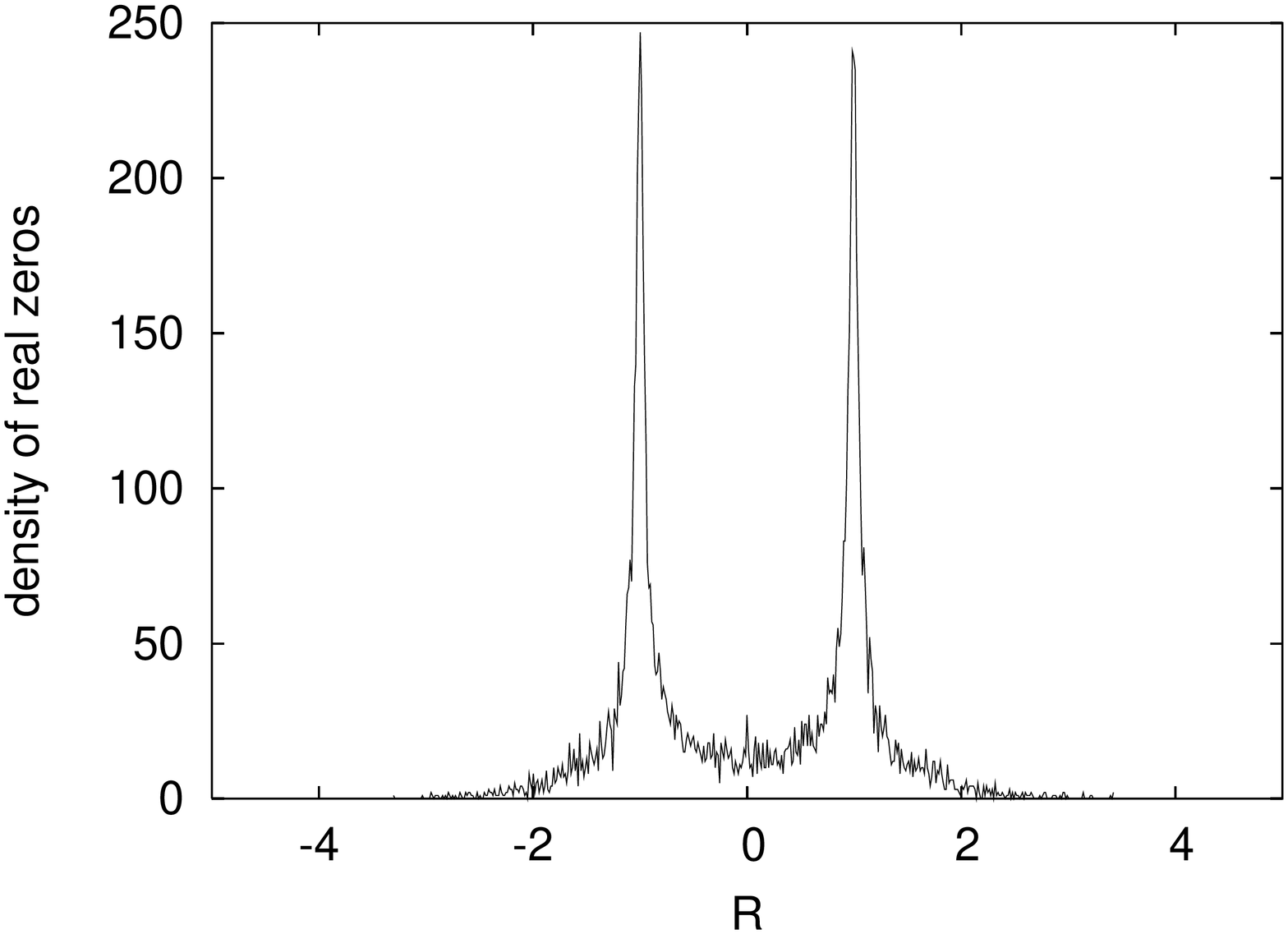, height=8.5cm, angle=270}
\end{center}
\caption{On the left is the theoretical real zero density for a $64$-degree
  polynomial with random coefficients drawn from normals with mean
  zero and unit variance.  On the right is the real zero density for a
  set of $3000$ companion matrices with $a_k$'s drawn from standard
  normals with mean zero and unit variance.}
\label{fig:zero_poly_density}
\end{figure}

\begin{figure}[tbp]
\begin{center}
\epsfig{file=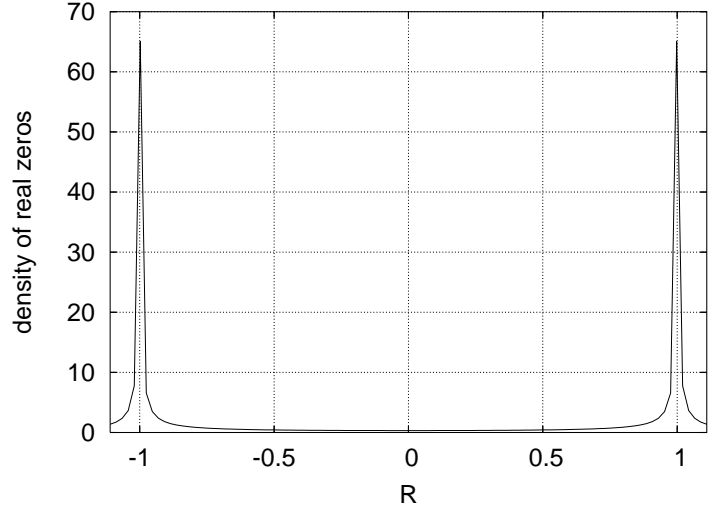, height=10cm, angle=270}
\end{center}
\caption{Theoretical density of real zeros for polynomials of degree
  $1024$ with coefficients drawn from a Gaussian distribution with
  mean zero and unit variance.  For $|x|>1.12$ the density is zero.}
\label{fig:zero_poly_density_highd}
\end{figure}

This seeming contradiction, a decreasing fraction of real zeros with a
constant fraction of bifurcations due to real eigenvalues, is rooted in the convergence of the density of real eigenvalues.  Figure
\ref{fig:zero_poly_density} shows both the empirical distribution
(of $1000$ polynomials) and
the theoretical density of real zeros for polynomials of degree $64$
with coefficients drawn from a Gaussian distribution with unit
variance.  The obvious spikes at $\pm 1$ are of interest, of course;
however, it is the tails that extend above and beyond $\pm 1$ that have
the most impact.  This feature can be highlighted at $d=1024$ if one
considers Fig. \ref{fig:zero_poly_density_highd}.  The point is that
the complex eigenvalues exist largely on the unit circle while the
real zeros have a non-zero measure set of zeros that have magnitude
greater than one.  The fraction of real bifurcations is largely
determined by the convergence of the density of real eigenvalues ---
and because the convergence has a non-zero density with magnitudes
greater than one, bifurcations due to real eigenvalues will persist to
very high-dimensional matrices or polynomials.  The fraction of real
bifurcations is constant because the ratio of the measure of \textit{complex}
eigenvalues with magnitude greater than one to the measure of \textit{real}
eigenvalues with magnitude greater than one is relatively constant
with increasing dimension.  


\subsection{Neural Networks} 
\label{sec:neural_net_numerical}

\begin{figure}[tbp]
\begin{center}
\epsfig{file=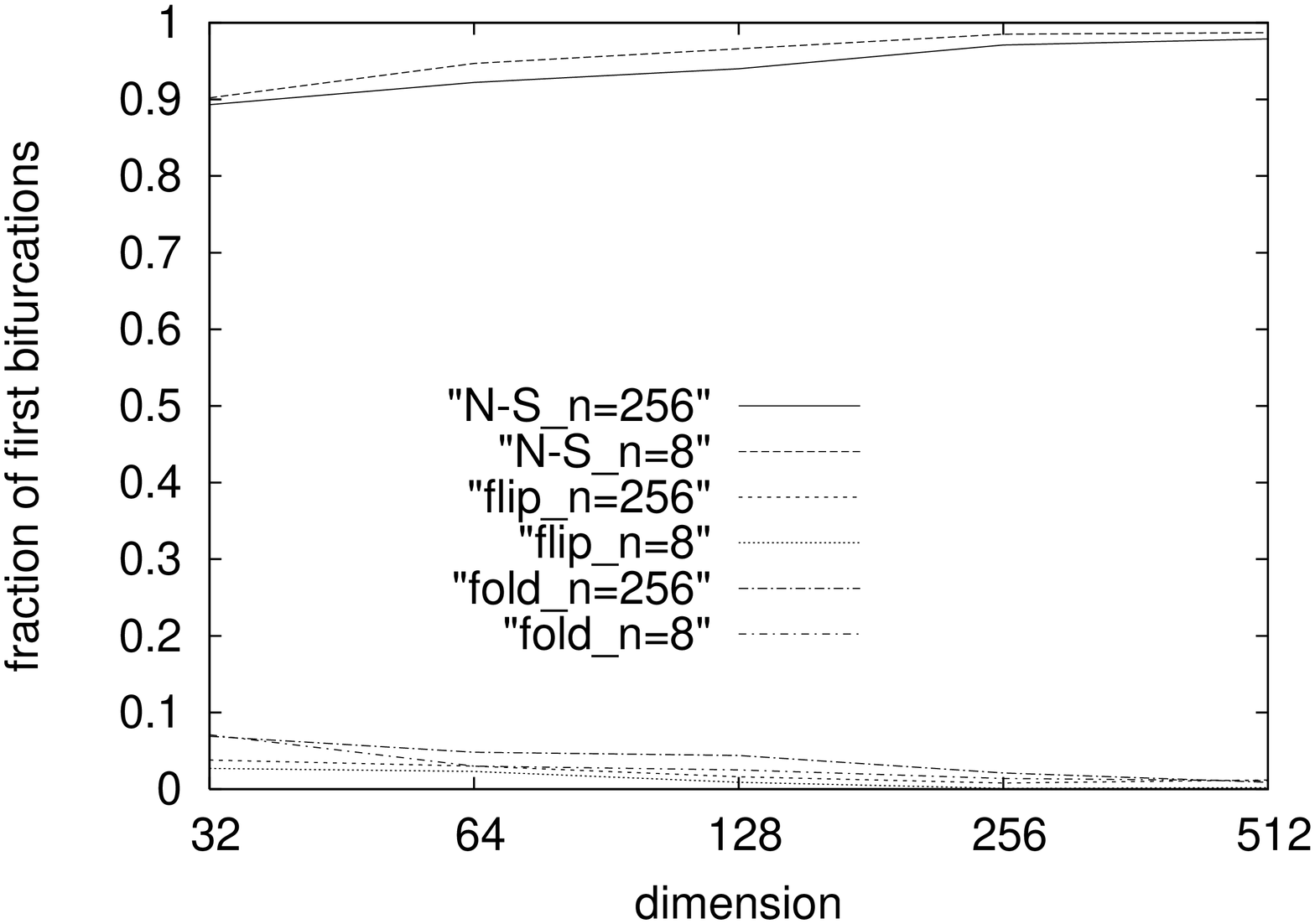, height=8.5cm, angle=270}
\epsfig{file=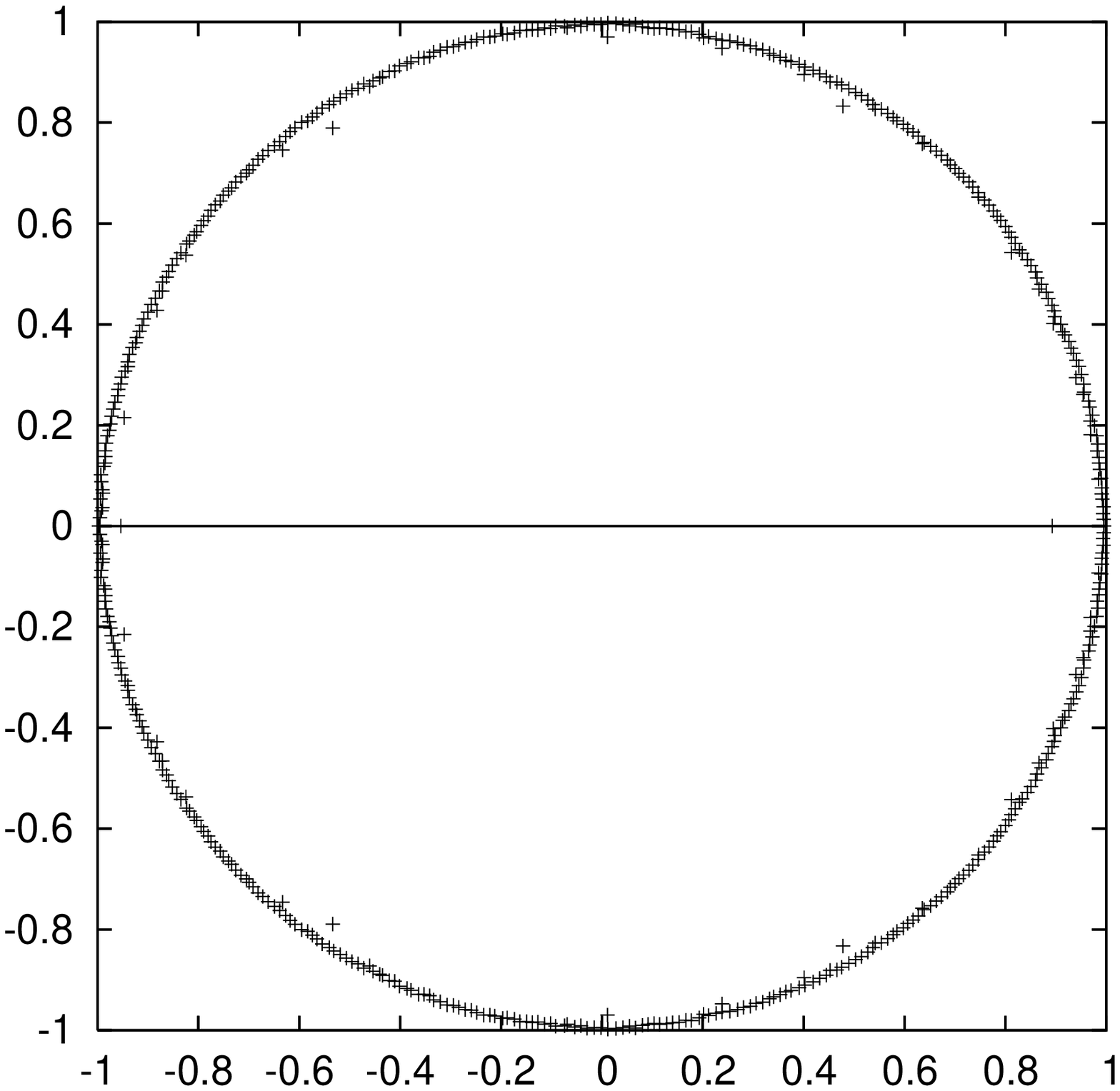, height=8.5cm, angle=270}
\end{center}
\caption{On the left, the observed probability of each bifurcation was
recorded for $1000$ neural networks for each $d$ along
with the fraction of eigenvalues that are real.  On the right is the
spectrum of eigenvalues in the complex plane that corresponds to a
single neural network with $n=256$ and $d=512$.}
\label{fig:nn_bif_fraction}
\end{figure}

\begin{figure}[tbp]
\begin{center}
\epsfig{file=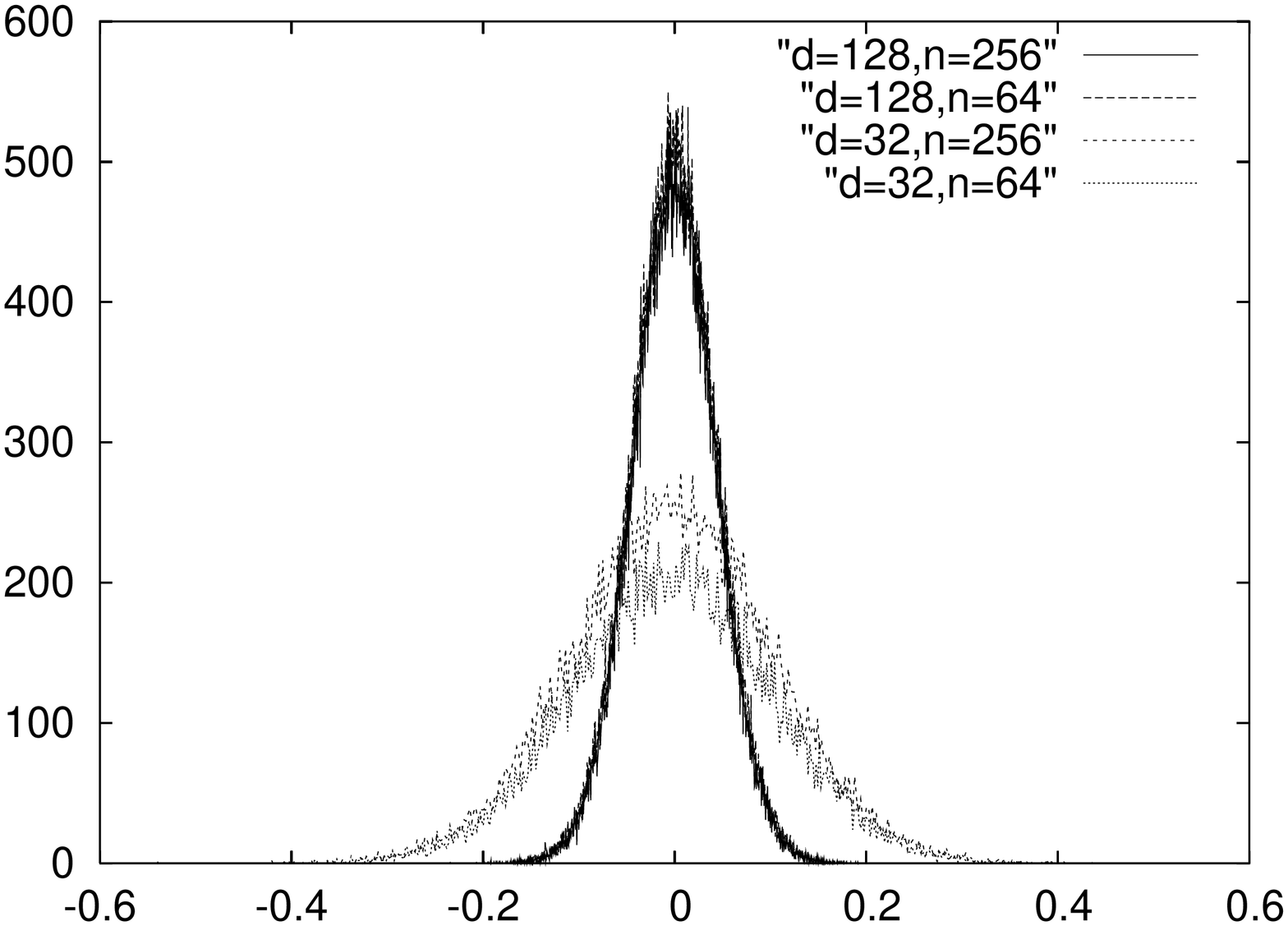, height=8.5cm, angle=270}
\epsfig{file=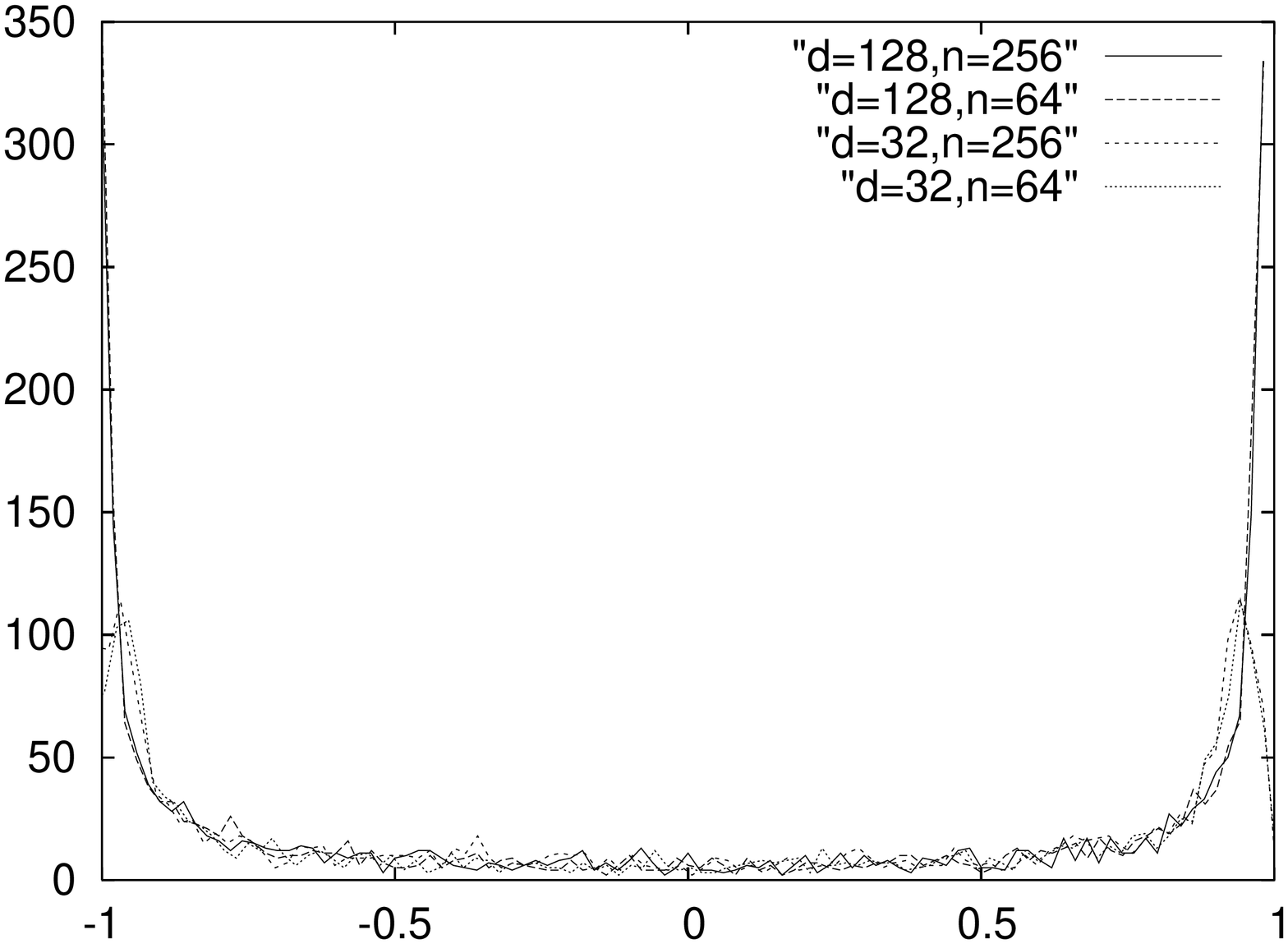, height=8.5cm, angle=270}
\end{center}
\caption{The plot on the right is of the distribution of $a_k$'s for
  $1000$ neural networks with $n=256, 64$ and $d=128, 32$.  The plot
  on the right is of the distribution of real eigenvalues along the
  real axis for the same set of neural networks.}
\label{fig:nn_ak_re}
\end{figure}

\begin{figure}[tbp]
\begin{center}
\epsfig{file=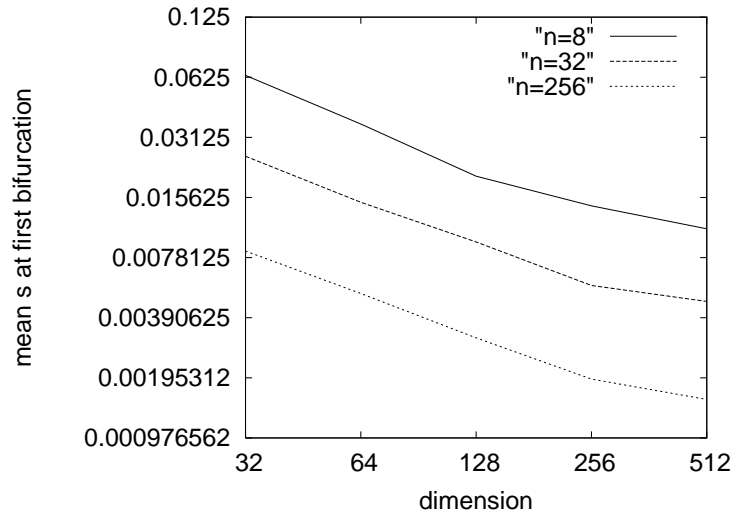, height=10cm, angle=270}
\end{center}
\caption{The mean location of $s$ at the first bifurcation for $n=8$,
  $n=32$, and $n=256$ for networks of dimensions varying from $32$ to $512$.}
\label{fig:nn_first_bif}
\end{figure}

In the case of neural networks, we do not have the luxury of having a formula
to guide our understanding of the empirical distribution of the real
eigenvalues or the $a_k$ values.  Compare Figs. \ref{fig:companion_bif_fraction}
and \ref{fig:nn_bif_fraction}.  In a preliminary comparison between companion matrices with Gaussian $a_k$'s and the
neural networks with the weight structure (defined in section
\ref{sec:construction}), the distribution of eigenvalues appears
nearly the same.  However, the fraction of
bifurcations of a fixed point that correspond to Naimark-Sacker type
in the neural networks tends towards
unity as the dimension tends toward infinity while the fraction of
Naimark-Sacker bifurcations in the companion matrices with Gaussian
$a_k$'s tends toward a constant value ($\sim 0.58$) as the dimension
is increased.  This difference can be rectified considering
Fig. \ref{fig:nn_ak_re} which depicts the distribution of real zeros
and the $a_k$'s for the neural networks.  With respect to the real
zeros, adding dimensions has very little effect on the interval
$(-0.9, 0.9)$.  However, near $\pm 1$, the real eigenvalues are
considerably more dense.  However, there do not exist the tails above
and below $\pm 1$ that are present in the companion matrices with
Gaussian $a_k$'s, which is enough to allow the fraction of
bifurcations due to real zeros to tend to zero as the dimension is
increased.  At first glance, increasing $d$ has a significant effect on the
variance $a_k$'s at the first bifurcation.  Because $s$ controls the variance of the $a_k$'s this effect is due
to the $s$ dependence of the first bifurcation.  The $s$ dependence can be understood by considering
Fig. \ref{fig:nn_first_bif} which characterizes the decrease in the
mean $s$ at the first bifurcation.  The decrease in the mean $s$-location of
the first bifurcation obeys the power law $\sim a d^{0.63}$ where $a$ depends on $n$.  Thus, with
increasing $d$, the variance of the $a_k$'s at the first bifurcation
point will decrease.  In general, the number of neurons has a
negligible effect on the distribution of real eigenvalues and only a
relatively minor effect on the peak of the $a_k$'s.




\section{Revisiting the conjecture}

The measure provided by Bai on random matrices --- any distribution
with a finite sixth moment --- is quite general; it was originally hoped
that such a measure would be enough to qualify probable bifurcations
from fixed points in a large set of dynamical systems.  As is now
clear, simply perturbing the first moment of the distribution, while
having no effects upon Bai's results, completely alters the probably
of a bifurcation.  In an intuitive sense, the difference corresponds
to dynamical systems whose most slowly contracting directions are
rotations versus dynamical systems for which a single, non-rotational
contraction dominates the slowly contracting dimension.  What we are
left with is a much more complicated picture.  It is likely, given our
numerical results, that corollary \ref{corollary:firstbifurcation} can
be generalized to distributions with finite sixth moment and zero
mean; however, we have not attempted to do so.  Conjecture
\ref{conjecture:firstbifurcation} is in part justified by the random
matrix results presented in section \ref{sec:numerical_cases}, but
clearly there is much room for a more complete numerical and
analytical study.  The apparent
distribution independence in the standard random matrix
case is not present in the companion matrix case.  Simply considering
the examples given by Edelman or the ones presented in the paper are
enough to demonstrate the existence of the diversity in the eigenvalue
spectrum with changes in the measure imposed on the $a_k$'s.  While the
analysis of the Gaussian $a_k$'s yields negative evidence for
conjecture \ref{conjecture:firstbifurcation_comp}, the neural network
analysis yields an example of a distribution of $a_k$'s that supports
conjecture \ref{conjecture:firstbifurcation_comp}.  The constraints that are required of a distribution of $a_k$'s that
will yield adherence to conjecture
\ref{conjecture:firstbifurcation_comp} are to the authors completely
unknown.  In preliminary studies, One property that seems important for
\ref{conjecture:firstbifurcation_comp} to be valid is bounded distributions.

Comparison of the full random matrix and companion matrix cases is
striking; the eigenvalue distributions for full random matrices are
distributed on the unit disc as opposed to the unit circle in the case
of companion matrices.  However, there is, via the neural network
approximation scheme, a very strong connection between the these two
situations that
has yet to be found.  Recalling the discussion in section
\ref{sec:approximation}, begin with a dynamical system 
\begin{equation}
F(x_{t-1}) = x_t = \epsilon A x_{t-1} 
\end{equation}
where $\epsilon \in R$ is small enough such that $F$ is a fixed point,
$A$ is a $d \times d$ real random matrix with
Gaussian elements with mean zero and unit variance.  There
exists a neural network $f$ of dimension $2d+1$ that can be trained on
the time-series generated by $F$ such that $f$ will have a spectrum
for which $d$ of it's eigenvalues will be identical (within the
desired numerical accuracy) as well as $d + 1$ free eigenvalues that
will have magnitude less than one.  Despite how the $d+1$ ``free''
eigenvalues are distributed, the remaining $d$ eigenvalues will have a
distribution that is not on the unit circle and thus significantly
different than any situations we have presented or know about.  This
connection yields insight into the distributions of $a_k$'s of companion matrices (and thus the
characteristic polynomial), as well as, forging a connection between standard dynamical systems and general
time-delay dynamics.  Finally, the neural networks provide an
opportunity for a connection between real-world systems and the
abstract dynamical models many in the field in dynamical systems study
via training, and an understanding how their weight distributions
affect their spectra.

\section{Beyond fixed points}
Extending this construction beyond bifurcations of fixed points to
the routes to chaos offers considerable problems.  There are two
basic approaches.  One involves reduction of bifurcations for periodic
orbits to bifurcations of fixed points in appropriately chosen
coordinates.  The other involves studying products of matrices of
derivatives of periodic orbits.  

Regarding the reduction of bifurcations of periodic orbits, one major problem
arises because bifurcations of periodic orbits must be understood well
enough to be approximated and reduced to analysis of fixed points.  To
see some of the currently open problems see \cite{kuzbook}.  Likewise,
see \cite{arrowsmithandplace} for nice explanations of the various approximation schemes.
Then the measures on the random matrices must be carried through the
various approximations.    

Taking products of random matrices might be a fruitful approach (see
\cite{ams_rmt} for results and techniques along these lines).  However,
linking them to periodic orbits might be difficult.  

Various authors have studied the routes to chaos computationally.  Utilizing the neural network framework discussed
here \cite{albersroutetochaosII} conclude that the most likely route to chaos is a
quasi-periodic one --- however these results are likely subject to the
measures imposed upon the weight matrices.  Likewise, Doyon
et. al. \cite{doy} \cite{doy2}, Cessac
et. al. \cite{cessac_mean_field_rtc}, and Sompolinsky et. al
\cite{Sompolinsky_nn} have arrived at similar conclusions in a variety
of circumstances.  Clearly
any results such as these will be subject to the same dependences on
the measures imposed on the weight matrices as were present in the
case of the bifurcation of a fixed point mentioned in the previous section.


\section{Final remarks}
In the context of a general dynamical system if (i) the Jacobian
of the fixed point can be identified with a full random matrix and
(ii) if the distribution of elements has zero mean, then, for the cases we
considered (uniform and Gaussian), as the dimension approaches
infinity, the probability of a Naimark-Sacker bifurcation approaches
unity.  However, the largest \textit{real} eigenvalue in the aforementioned
circumstances scales linearly with the mean of the distribution of the
elements of the matrix.  Therefore, if the mean is large, the
most probable bifurcation will be due to a real eigenvalue (e.g. a flip or
a fold bifurcation), regardless of the dimension.  An analytical
understanding of linear scaling of the largest real eigenvalue with
the mean of the distribution is unknown.  Aside from the effect of the
mean on the largest real eigenvalue, the result of the probability of a Naimark-Sacker bifurcation
increasing with dimension is quite independent of the distribution.
This is because, for nearly all distributions of elements, real random
matrices have a spectrum that converges to uniformity on the unit disc
as the dimension of the matrix goes to infinity. 

In the context of time-delay dynamical systems, the story is
different.  In this case, the Jacobian forms a companion matrix.  If
the companion matrix has elements drawn from a Gaussian distribution,
the probability of a Naimark-Sacker bifurcation saturates at $\sim 58$
percent as the dimension goes to infinity.  This is due to tails in
the distribution of real zeros.  Numerical results with time-delay,
feed-forward neural networks however, behave very differently.  For
neural networks, as the dimension goes to infinity, the probably of a
Naimark-Sacker bifurcation goes to unity.  For companion matrices, the
first bifurcation probability is highly dependent on the distribution
of the elements of the matrix.  In general, the spectrum of a
time-delay dynamical system lies on the unit \textit{circle} and the
real line, not the entire unit disc as in the case with full
matrices.  However, the distriution of real zeros can vary 
significantly from distribution to distribution.  Providing a link between the
time-delay case and the standard dynamical systems framework via
neural network training is
suggested as future work.

\section{Acknowledgments}
We thank R. A. Bayliss, W. D. Dechert, W. Brock, J. Albers, J. Thompson,
J. Jost, I. Dobson, C. McTague, and J. C. Crutchfield for helpful discussions.
This work was partially supported at the Santa Fe Institute under
the Networks Dynamics Program funded by the Intel Corporation and under the
Computation, Dynamics, and Inference Program via SFI's core grants from the
National Science and MacArthur Foundations.  Computing was done
primarily on the Max Planck Institute for Mathematics
in the Sciences Beowulf cluster, however significant computations were
carried out on the U.C. Davis Computational Science Beowulf cluster
and the Santa Fe Institute Beowulf cluster.


\bibliography{dstexts,partialhyperbolicity,lyapunovexponents,neuralnetworks,nilpotency,topology,analysis,structuralstability,computation,me,physics,bifurcationtheory,probability,lorenz,srb,unimodal,geometry}

\end{document}